\documentclass[epsfig,12pt]{article}
\usepackage{makeidx}
\usepackage{amsmath}
\usepackage{amsfonts}
\usepackage{amssymb}
\usepackage{graphicx}

\input epsf.sty
\newtheorem{theorem}{Theorem}
\newtheorem{acknowledgement}[theorem]{Acknowledgement}

\makeatletter \@addtoreset{equation}{section}
\renewcommand{\theequation}{\thesection.\arabic{equation}}
\input{tcilatex}
\begin{document}

\title{\textbf{Extremal Black Attractors in} \emph{8D} \textbf{Maximal
Supergravity}}
\author{L.B. Drissi$^{1}$, F.Z Hassani$^{2}$, H. Jehjouh$^{2}$, E.H Saidi$%
^{1,2}$  \\
{\small a. INANOTECH, Institute of Nanomaterials and Nanotechnology, Rabat,
Morocco,}\\
{\small b. LPHE, Modelisation et Simulation, Facult\'{e} des Sciences Rabat,
Morocco}}
\maketitle

\begin{abstract}
Motivated by the new higher D-supergravity solutions on intersecting
attractors obtained by \emph{Ferrara et al.} in [\textrm{\
Phys.Rev.D79:065031-2009}], we focus in this paper on \emph{8D }maximal
supergravity with moduli space $\frac{SL\left( 3,R\right) }{SO\left(
3\right) }\times \frac{SL\left( 2,R\right) }{SO\left( 2\right) }$ and study
explicitly the attractor mechanism for various configurations of extremal
black p- branes (anti-branes) with the typical near horizon geometries $%
AdS_{p+2}$ $\times $ $S^{m}$ $\times $ $T^{6-p-m}$ and $p=0,1,2,3,4;$ $2\leq
m\leq 6$.\ Interpretations in terms of wrapped M2 and M5 branes of the 11D
M-theory on 3-torus are also given.\newline
\end{abstract}


\section{Introduction}

Since its discovery by \emph{D.Z Freedman, P.van Nieuwenhuizen and S.Ferrara}
at the mid of seventieth of the last century \textrm{\cite{1A,1B}}, the
properties of supergravity theories, based on the gauging of
Haag-Lopozansky-Sohinus (HLS) supersymmetry \textrm{\cite{2}}, have been
intensively studied in four and higher dimensions; for reviews see \textrm{%
\cite{3A,3B,3C}}. These studies allowed more insight into supersymmetric
gauge theories in diverse dimensions and led to the obtention of
superstrings \textrm{\cite{4}} containing the various \emph{4D} and higher
supergravities as Planck scale limits of \emph{10D} superstrings and \emph{%
11D} M theory compactifications \textrm{\cite{4B}}. Besides usual
properties, supergravity in higher dimensions have moreover specific
features; in particular they need a graded Lie algebraic structure going
beyond the LHS superalgebra by implementing exotic \textquotedblleft
central\textquotedblright\ charges $\mathcal{Z}_{{\scriptsize [}\mu
_{1}...\mu _{p}{\scriptsize ]}}\equiv \mathcal{Z}_{p}$ that transform in non
trivial representations of $SO\left( 1,D-1\right) $ space time symmetry 
\textrm{\cite{5A,5B}}. They also have $\left( p+1\right) $- form gauge
fields $\mathcal{A}_{p+1}\equiv \mathcal{A}_{{\scriptsize [}\mu _{1}...\mu
_{p+1}{\scriptsize ]}}$ in addition to the graviton $\mathcal{G}_{\mu \nu }$%
, the usual 1-form gauge fields $\mathcal{A}_{\mu }$, scalars $\left\{ \phi
^{I}\right\} $ and their supersymmetric partners $\psi _{\mu }^{\alpha },$ $%
\chi ^{\alpha }$. These central charges $\mathcal{Z}_{p}$ and gauge fields $%
\mathcal{A}_{p+1}$ play, like in the case of \emph{4D} black holes \textrm{%
\cite{5C,5D,5E}}, a crucial role in dealing with static, asymptotically flat
and spherically symmetric extremal black p-brane solutions living in higher
dimensions. In this regards and following the first original works \textrm{%
\cite{6A,6B,6C,6D,6E,6F} }and subsequent ones led by S.Ferrara and
collaborators \textrm{\cite{6G,6H} }and refs therein, growing attention has
been devoted to the study of the black hole solutions in various dimensions
and their attractor mechanism taking into account p-branes carrying non
trivial magnetic $p^{\Lambda }$ and electric charges $q_{\Lambda }$ of the $%
\left( p+2\right) $-form gauge field strengths $\mathcal{F}_{p+2}^{\Lambda }$
and their magnetic duals $\mathcal{\tilde{F}}_{{\scriptsize D-p-2|}\Lambda }$
\textrm{\cite{7A}-\cite{7Z}}. The attractor equations are obtained by
minimization of the effective potential $\mathcal{V}_{eff}\left( \phi
\right) $ induced by the kinetic energies of the gauge field strengths of
the supergravity theory. The minima of the effective potential, solving the
conditions $\partial _{I}\mathcal{V}_{eff}=0$, $\det \left( \partial
_{I}\partial _{J}\mathcal{V}_{eff}\right) >0$, $\delta \mathcal{V}_{eff}=0$,
determine the values of the scalars at the horizon in terms of the black
brane charges $p^{\Lambda }$ and $q_{\Lambda }$.\newline
Motivated by the new solutions on higher dimensional intersecting attractors
recently obtained in \textrm{\cite{8A}}, we focus in this paper on maximal
supergravity in \emph{8D} with moduli space $\left[ SL\left( 3,R\right)
/SO\left( 3\right) \right] \times \left[ SL\left( 2,R\right) /SO\left(
2\right) \right] $ and study \emph{explicitly} the attractor mechanism for
various configurations of black p- branes and anti-branes living in \emph{8D 
}and having the typical near horizon geometries $AdS_{p+2}$ $\times $ $S^{m}$
$\times $ $T^{6-p-m}$, with $0\leq p\leq 4,$ and $2\leq p+m\leq 6$. We also
complete some partial results of \cite{8A}; in particular the strand on the
black dyonic membrane and the dual black attractor pairs string/(anti)
3-branes, holes/(anti) 4-branes.\newline
The presentation is as follows: \emph{In section 2}, we first study the 
\emph{8D} $\mathcal{N}=\left( 2,2\right) $ supersymmetric algebra in
presence of p-branes; then we consider the embedding of this non chiral
supersymmetric field theory into \emph{11D} M theory on the 3-torus. This is
useful for learning the group theoretic representations in which the gauge
and scalar fields transform. \emph{In section 3}, we study the attractor eqs
for the black branes in \emph{8D}. We first consider an unconstrained
parametrization of the moduli space, then we study the total effective
potential and we derive the general form of the attractor eqs depending on
the values of the Maurer Cartan 1-forms. \emph{In section 4}, we study the
solutions for the attractors equations. We study the explicit solutions for
the dyonic membrane; actually, this completes the analysis done in \textrm{%
\cite{8A}}. Then we consider the general solutions for case of black strings
and black 3-branes. This study extends directly to the case of black holes /
black 4-branes; which is omitted. \emph{In section 5}, we make an explicit
study of the intersecting attractors in \emph{8D} by using the approach of 
\textrm{\cite{8A}}. \emph{In section 6} we give our conclusion and in \emph{%
in section 7}, we give an appendix on useful properties on the algebras of
spinors in 8D space time.

\section{Z- charges in \emph{8D }$\mathcal{N}=\left( 2,2\right) $\emph{\ }%
supergravity}

We begin by studying maximal supersymmetry in eight dimensional space time
with p-branes. Then, we consider the embedding of \emph{8D }$\mathcal{N}%
=\left( 2,2\right) $\emph{\ }supergravity into the \emph{11D }M-theory
compactification on the 3-torus $T^{3}$. Configurations based on M2 and M5
branes wrapping various cycles of $T^{3}$ are also considered in connection
with black p-branes in \emph{8D}.

\subsection{$\mathcal{N}=\left( 2,2\right) $\emph{\ }superalgebra with branes%
}

In eight dimensions, non chiral $\mathcal{N}=\left( 2,2\right) $
supersymmetry has \emph{32} conserved supersymmetric charges given by the 
\emph{8D} fermionic generators, 
\begin{equation}
\begin{tabular}{lll}
{\small generators} & : & $SO\left( 1,7\right) \times SU\left( 2\right)
\times U\left( 1\right) $ \\ 
$\ \ \ \ Q_{\alpha }^{+A}$ & $\sim $ & $\ \ \ \ \ \ \ \ \ \left(
8_{s},2\right) _{+}$ \\ 
$\ \ \ \ Q_{\dot{\alpha}A}^{-}$ & $\sim $ & $\ \ \ \ \ \ \ \ \ \left( 8_{c},%
\bar{2}\right) _{-}$%
\end{tabular}
\label{82}
\end{equation}%
In addition to the $SO\left( 1,7\right) $ space time, we also have an extra $%
U\left( 2\right) =U\left( 1\right) \times SU\left( 2\right) $ invariance;
this is an automorphism symmetry group with the $U\left( 1\right) $ factor
capturing the $\pm $ chirality charges of the Weyl spinors in \emph{8D} and
the $SU\left( 2\right) $ rotating the two supercharges $Q^{\pm A}$.%
\begin{equation}
\begin{tabular}{llll}
$Q^{+A}\longrightarrow e^{i\frac{q}{2}\theta }\left( U_{B}^{A}Q^{+B}\right) $
& , & $Q_{A}^{-}\longrightarrow Q_{B}^{-}\left( U^{\dagger }\right)
_{A}^{B}e^{-i\frac{q}{2}\theta }$ & ,%
\end{tabular}%
\end{equation}%
with $U\left( 1\right) $ charge $q=1$ and $U$ a unimodular $2\times 2$
unitary matrix. This $SU\left( 2\right) $ automorphism symmetry has also an
interpretation as an internal symmetry in terms of embedding $\mathcal{N}%
=\left( 2,2\right) $ \emph{8D} supergravity in the \emph{11D} M-theory
compactification on $T^{3}$. Under the reduction from \emph{11D} down to 
\emph{8D}, the $SO\left( 1,10\right) $ Lorentz group at each point of space
time $\mathcal{M}_{11}$ gets broken down to $SO\left( 1,7\right) \times
SO\left( 3\right) $ where the internal $SO\left( 3\right) $ orthogonal group
is thought of in terms of the covering $SU\left( 2\right) $ symmetry. 
\newline
To get the general structure of the supersymmetric Lie algebra satisfied by
the $Q_{\alpha }^{+A}$ and $Q_{\dot{\alpha}A}^{-}$ operators, we use results
on the tensor products of $SO\left( 1,7\right) \times SU\left( 2\right)
\times U\left( 1\right) $ representations; in particular the $SO\left(
1,7\right) $ ones, 
\begin{equation}
\begin{tabular}{llll}
$8_{i}\times 8_{i}$ & $=$ & $1+28+35_{i}$ & , \\ 
$8_{i}\times 8_{j}$ & $=$ & $8_{k}+56_{k}$ & ,%
\end{tabular}%
\end{equation}%
with $i,j,k$ cyclic and where $8_{i}$ stand for \emph{8}$_{s}$, \emph{8}$%
_{c} $, \emph{8}$_{v}$ describing the basic eight dimensional
representations of $SO\left( 1,7\right) $. For the case of spinor doublets
of eqs(\ref{82}), we have the decompositions,%
\begin{equation}
\begin{tabular}{llll}
$\left( 8_{s},2\right) \otimes \left( 8_{s},2\right) $ & $=$ & $\left(
1,4\right) \oplus \left( 28,4\right) \oplus \left( 35_{s},4\right) $ &  \\ 
$\left( 8_{s},2\right) \otimes \left( 8_{c},\bar{2}\right) $ & $=$ & $\left(
8_{v},4\right) \oplus \left( 56_{v},4\right) $ &  \\ 
$\left( 8_{c},\bar{2}\right) \otimes \left( 8_{c},\bar{2}\right) $ & $=$ & $%
\left( 1,4\right) \oplus \left( 28,4\right) \oplus \left( 35_{c},4\right) $
& 
\end{tabular}
\label{8V}
\end{equation}%
with the complex $\left( 1+35\right) $ associated with the $\frac{8\times 9}{%
2}$ symmetric part of the product and the complex $28=\frac{8\times 7}{2}$
with the antisymmetric component. Using these relations, the general form of
the anticommutation relations between the fermionic generators $Q_{\alpha
}^{+A}$ and $Q_{\dot{\alpha}A}^{-}$ may be written as follows,%
\begin{equation}
\begin{tabular}{lll}
$\left\{ Q_{\gamma }^{+A},Q_{\dot{\delta}B}^{-}\right\} $ & $=$ & $\Gamma
_{\gamma \dot{\delta}\ }^{\mu }\delta _{B}^{A}P_{\mu }+Z_{\gamma \dot{\delta}%
|B}^{{\scriptsize 0}{\small A}}$ \\ 
$\left\{ Q_{\alpha }^{+A},Q_{\beta }^{+B}\right\} $ & $=$ & $Z_{\alpha \beta
}^{++AB}$ \\ 
$\left\{ Q_{\dot{\gamma}A}^{-},Q_{\dot{\delta}B}^{-}\right\} $ & $=$ & $Z_{%
\dot{\gamma}\dot{\delta}|AB}^{--}$%
\end{tabular}
\label{SS}
\end{equation}%
where in addition to the usual terms $\Gamma _{\gamma \dot{\delta}\ }^{\mu
}\delta _{B}^{A}P_{\mu }$, we have moreover other charge operators
transforming into non trivial representations of $SO\left( 1,7\right) $.
These operators have the following expansion properties%
\begin{equation}
\begin{tabular}{ll}
$Z_{\gamma \dot{\delta}|B}^{0{\small A}}$ & $=\varepsilon _{BC}\Gamma
_{\gamma \dot{\delta}\ }^{\mu }Z_{\mu }^{0\left( AC\right) }+\varepsilon
_{BC}\Gamma _{\gamma \dot{\delta}\ }^{\mu \nu \rho }Z_{\mu \nu \rho }^{0AC}$
\\ 
$Z_{\alpha \beta }^{++AB}$ & $=\delta _{\alpha \beta }Z^{++\left( AB\right)
}+\varepsilon ^{AB}\Gamma _{\alpha \beta }^{\mu \nu }Z_{\mu \nu
}^{++}+\Gamma _{\alpha \beta }^{\mu \nu \rho \lambda }Z_{\mu \nu \rho
\lambda }^{++\left( AB\right) }$ \\ 
$Z_{\dot{\gamma}\dot{\delta}|AB}^{--}$ & $=\delta _{\dot{\gamma}\dot{\delta}%
}Z_{\left( AB\right) }^{--}+\varepsilon _{{\scriptsize AB}}\Gamma _{\dot{%
\gamma}\dot{\delta}}^{\mu \nu }Z_{\mu \nu }^{--}+\varepsilon _{{\scriptsize %
AC}}\varepsilon _{{\scriptsize BD}}\Gamma _{\gamma \delta }^{\mu \nu \rho
\lambda }Z_{\mu \nu \rho \lambda }^{--\left( CD\right) }$ \\ 
& 
\end{tabular}%
\end{equation}%
where anti-symmetrization with respect to the space time indices is
understood; see also appendix for more details on $\Gamma $-\ matrices.
Obviously, the charge operators $Z_{\gamma \dot{\delta}|B}^{{\scriptsize 0}%
{\small A}}$, $Z_{\alpha \beta }^{++AB}$ and $Z_{\dot{\gamma}\dot{\delta}%
|AB}^{--}$ are bosonic and generally take non zero values; they transform
non trivially under $SO\left( 1,7\right) $ rotations and obey commutation
relations \textrm{\cite{5A,5B}}, that are obtained as usual by solving the
graded Jacobi identities. Let us comment much more these objects as they are
crucial in studying black p-branes. The operators $Z_{\alpha \beta }^{++AB}$
are complex and correspond to taking the symmetric part of the following
tensor product relation,%
\begin{equation}
\begin{tabular}{llll}
$\left( 8_{s},2\right) _{{\scriptsize +}}\otimes \left( 8_{s},2\right) _{%
{\scriptsize +}}$ & $=$ & $\left( 1,1\right) _{{\scriptsize ++}}\oplus
\left( 28,1\right) _{{\scriptsize ++}}\oplus \left( 35_{s},1\right) _{%
{\scriptsize ++}}\oplus $ &  \\ 
&  & $\left( 1,3\right) _{{\scriptsize ++}}\oplus \left( 28,3\right) _{%
{\scriptsize ++}}\oplus \left( 35_{s},3\right) _{{\scriptsize ++}}$ & 
\end{tabular}%
\end{equation}%
where the $U\left( 1\right) $ charges are exhibited as sub-indices. This
decomposition leads to the identifications%
\begin{equation}
\begin{tabular}{lllll}
$Z^{++\left( AB\right) }\sim \left( 1,3\right) _{{\scriptsize ++}}$ & , & $%
Z_{\mu \nu }^{++}\sim \left( 28,1\right) _{{\scriptsize ++}}$ & , & $Z_{\mu
\nu \rho \lambda }^{++\left( AB\right) }\sim \left( 35_{s},3\right) _{%
{\scriptsize ++}}$%
\end{tabular}
\label{bt}
\end{equation}%
and shows that there various kinds of Z-charge operators capturing a priori
different information of $\mathcal{N}=\left( 2,2\right) $ supersymmetric
theory in \emph{8D}. This is in fact what happens as we will see throughout
this study. Notice that similar relations are valid for $\left(
8_{c},2\right) _{{\scriptsize -}}\otimes \left( 8_{c},2\right) _{%
{\scriptsize -}}$; they are just the complex conjugates of the above ones.
We also have%
\begin{equation}
\begin{tabular}{lll}
$\left( 8_{s},2\right) _{{\scriptsize +}}\otimes \left( 8_{c},\bar{2}\right)
_{{\scriptsize -}}$ & $=$ & $\left( 8_{v},1\right) _{{\scriptsize 0}}\oplus
\left( 56_{v},1\right) _{{\scriptsize 0}}\oplus \left( 8_{v},3\right) _{%
{\scriptsize 0}}\oplus \left( 56_{v},3\right) _{{\scriptsize 0}}$%
\end{tabular}%
\end{equation}%
with the correspondence%
\begin{equation}
\begin{tabular}{lllll}
$Z_{\mu }^{\left( AB\right) }\sim \left( 8_{v},3\right) _{{\scriptsize 0}}$
& , & $Z_{\mu \nu \rho }^{0}\sim \left( 56_{v},1\right) _{{\scriptsize 0}}$
& , & $Z_{\mu \nu \rho }^{\left( AB\right) }\sim \left( 56_{v},3\right) _{%
{\scriptsize 0}}$%
\end{tabular}
\label{bi}
\end{equation}%
From this analysis, we learn amongst others that these bosonic $Z$-
generators appearing in the supersymmetric algebra (\ref{SS}) exhibit a set
of remarkable properties; in particular the two following:\newline
(\textbf{a}) Like the other generators of the superalgebra (\ref{SS}), the
Z's are generally charged under the internal $SU\left( 2\right) \times
U\left( 1\right) $ automorphism symmetry and $SO\left( 1,7\right) $
invariance since, in addition to the quantum numbers $\left( A,\pm q\right) $%
, they also carry space time indices in the antisymmetric representations.
The last property allows to associate to each $Z_{\mu _{1}...\mu _{p}}$
operator the space time p-form operator density 
\begin{equation}
\mathcal{Z}_{p}=\frac{1}{p!}dx^{\mu _{1}}{\scriptsize \wedge }...%
{\scriptsize \wedge }dx^{\mu _{p}}Z_{\mu _{1}...\mu _{p}},
\end{equation}%
together with the $\mathcal{J}\left( M_{p}\right) \equiv \mathcal{J}_{p}$
invariant, $\mathcal{J}_{p}=\int\nolimits_{M_{p}}\mathcal{Z}_{p}$, where $%
M_{p}$ is a p-dimensional space time submanifold which may be thought of as
the world volume of a p-brane. Using eqs(\ref{bt}-\ref{bi}), it follows that
in \emph{8D} maximal supergravity, we have for the complex $SU\left(
2\right) $ singlets $Z_{\mu \nu }^{\pm \pm }$, the 2-form operators%
\begin{equation}
\begin{tabular}{llll}
$\mathcal{Z}_{2}=\frac{1}{2}dx^{\mu }{\scriptsize \wedge }dx^{\nu }\text{ }%
Z_{\mu \nu }$ & , & $\mathcal{J}_{2}=\int\nolimits_{M_{2}}\mathcal{Z}_{2}$ & 
,%
\end{tabular}%
\end{equation}%
and the p-forms,%
\begin{equation}
\begin{tabular}{llll}
$\mathcal{Z}_{p}^{\left( AB\right) }=\frac{1}{p!}dx^{\mu _{1}}{\scriptsize %
\wedge }...{\scriptsize \wedge }dx^{\mu _{p}}\text{ }Z_{\mu _{1}...\mu
_{p}}^{\left( AB\right) }$ & , & $\mathcal{J}_{p}^{\left( AB\right)
}=\int\nolimits_{M_{p}}\mathcal{Z}_{p}^{\left( AB\right) }$ & ,%
\end{tabular}%
\end{equation}%
for $p=0,1,3,4$ for the $SU\left( 2\right) $ triplets.\newline
(\textbf{b}) From eqs(\ref{bt}-\ref{bi}), we learn as well that the $Z_{\mu
_{1}...\mu _{p}}$ operators have at most four space time indices. This
property allows to give them an interpretation in terms of fluxes of gauge
fields in \emph{8D} supergravity. Indeed, using the usual relations $%
m=\int\nolimits_{S^{2}}\mathcal{F}_{2}$ and $e=\int\nolimits_{S^{2}}\mathcal{%
\tilde{F}}_{2}$ giving the magnetic and electric charges of particles
coupled to \emph{4D}\ Maxwell gauge fields and thinking about the $\mathcal{Z%
}_{p}$'s in the same manner, we end with the following relations 
\begin{equation}
\begin{tabular}{lll}
$\mathcal{Z}_{p}=\int_{S^{2}}\mathcal{F}_{p+2}$ & , & $\mathcal{\tilde{Z}}%
_{4-p}=\int_{S^{2}}\mathcal{\tilde{F}}_{6-p}$ \\ 
$\mathcal{J}_{p}=\int_{M_{p+{\small 2}}}\mathcal{F}_{p+2}$ & , & $\mathcal{%
\tilde{J}}_{4-p}=\int_{\tilde{M}_{6-p}}\mathcal{\tilde{F}}_{6-p}$%
\end{tabular}%
\end{equation}%
teaching us that the $\mathcal{Z}_{p}$'s describe precisely charges of
p-branes that couple to the \emph{8D} supergravity $\left( p+1\right) $-
form gauge fields $\mathcal{A}_{p+1}$ with the field strengths $\mathcal{F}%
_{p+2}$ and their magnetic duals $\mathcal{\tilde{F}}_{6-p}$. In these
relations, the spaces $M_{p+{\small 2}}\sim M_{p}\times S^{2}$ and $\tilde{M}%
_{6-p}\sim \tilde{M}_{4-p}\times \tilde{S}^{2}$ are dual sub-manifolds of
the \emph{8D}\ space time $\mathcal{M}_{8}$ with the typical fibration,%
\begin{equation}
\begin{tabular}{llll}
$\tilde{M}_{6-p}$ & $\longrightarrow $ & $\mathcal{M}_{8}$ &  \\ 
&  & $\downarrow \pi _{p+2}$ &  \\ 
&  & $M_{p+{\small 2}}$ & 
\end{tabular}
\label{16}
\end{equation}%
where $M_{p+{\small 2}}$ is thought as the $\left( p+2\right) $- dimensional
base sub-manifold and $\tilde{M}_{6-p}$ as its $\left( 6-p\right) $-
dimensional fiber. For later use notice also that p-branes and their $\left(
4-p\right) $- duals extend along the dimensions of the respective $M_{p}$
and $\tilde{M}_{4-p}$ base sub-manifolds of $M_{p+2}$ and $\tilde{M}_{4-p}$, 
\begin{equation}
\begin{tabular}{lllllllllll}
$\ \ \ \ \ S^{2}$ & $\longrightarrow $ & $M_{p+2}$ &  &  &  &  &  & $\ \ \ \
\ \tilde{S}^{2}$ & $\longrightarrow $ & $\tilde{M}_{6-p}$ \\ 
&  & $\downarrow \pi _{{\small p}}$ &  &  & , &  &  &  &  & $\downarrow 
\tilde{\pi}_{{\small 4-p}}$ \\ 
\multicolumn{2}{l}{\ \ \ \ {\small p-branes}} & $M_{p}$ &  &  &  &  &  & 
\multicolumn{2}{l}{\ \ \ {\small 4-p branes}} & $\tilde{M}_{4-p}$%
\end{tabular}
\label{217}
\end{equation}%
Below we study some aspects on gauge fields in \emph{8D} supergravity; in
particular the gauge field content, the connection with p-branes and the
embedding in M-theory compactification the 3-torus.

\subsection{Embedding \emph{8D} $\mathcal{N}=\left( 2,2\right) $
supergravity in M-theory}

The massless spectrum of the supergravity limit of M-theory has, besides the 
\emph{11D} field metric $\mathcal{G}_{MN}^{\left( {\scriptsize 11D}\right) }$%
, an antisymmetric gauge 3-form $\mathcal{C}_{MNP}^{\left( {\scriptsize 11D}%
\right) }$ that couples to M2 brane as well as fermionic partners. Under
compactification of M-theory on the 3-torus, the M2 and M5 get wrapped and
the fields $\mathcal{G}_{MN}^{\left( {\scriptsize 11D}\right) }$ and $%
\mathcal{C}_{MNP}^{\left( {\scriptsize 11D}\right) }$ get reduced to, 
\begin{equation}
\begin{tabular}{llll}
$\mathcal{G}_{\mu \nu }^{\left( {\scriptsize 8D}\right) }$ & , & $\mathcal{C}%
_{\mu \nu \rho }^{\left( {\scriptsize 8D}\right) }$ & ,%
\end{tabular}%
\end{equation}%
together with the following \emph{8D }bosonic fields namely%
\begin{equation}
\begin{tabular}{llllll}
$\mathcal{B}_{\mu \nu }^{a}$ & , & $\mathcal{A}_{\mu }^{ai}$ & , & $\phi
^{1},...,\phi ^{7}$ & ,%
\end{tabular}%
\end{equation}%
where the indices $a=1,2,3$ and $i=1,2$. The field $\mathcal{C}_{\mu \nu
\rho }^{\left( {\scriptsize 8D}\right) }$ couples to the membrane M2 living
in \emph{8D }as in eqs(\ref{16}-\ref{217}), the three $\mathcal{B}_{\mu \nu
}^{a}$'s couple to the three kinds of strings obtained by wrapping the M2
brane on each $S^{1}$ cycle of $T^{3}$ as given below,%
\begin{equation}
\begin{tabular}{l}
$%
\begin{tabular}{l|l|l|l|l|l|l|l|l||lll|}
\cline{2-12}
& \multicolumn{8}{|l}{$\ \ \ \ \ \mathcal{M}_{8}$ {\small space time}} & 
\multicolumn{3}{||l|}{\ \ \ $\ \ \ \ \ \ \ \ T^{3}$} \\ \cline{2-12}
& 0 & 1 & 2 & 3 & 4 & 5 & 6 & 7 & 8 & 9 & 10 \\ \hline
\multicolumn{1}{|l|}{M2} & $\times $ & $\times $ & $\mathbf{\circ }$ & $%
\mathbf{\circ }$ & $\mathbf{\circ }$ & $\mathbf{\circ }$ & $\mathbf{\circ }$
& $\mathbf{\circ }$ & $\left\{ 
\begin{array}{c}
\mathbf{\times } \\ 
\circ \\ 
\circ%
\end{array}%
\right. $ & $\left. 
\begin{array}{c}
\mathbf{\circ } \\ 
\times \\ 
\circ%
\end{array}%
\right. $ & $\left. 
\begin{array}{c}
\mathbf{\circ } \\ 
\circ \\ 
\times%
\end{array}%
\right. $ \\ \hline
\multicolumn{1}{|l|}{M5} & $\mathbf{\circ }$ & $\mathbf{\circ }$ & $\mathbf{%
\circ }$ & $\mathbf{\circ }$ & $\times $ & $\times $ & $\times $ & $\times $
& $\left\{ 
\begin{array}{c}
\mathbf{\circ } \\ 
\times \\ 
\times%
\end{array}%
\right. $ & $\left. 
\begin{array}{c}
\mathbf{\times } \\ 
\circ \\ 
\times%
\end{array}%
\right. $ & $\left. 
\begin{array}{c}
\mathbf{\times } \\ 
\times \\ 
\circ%
\end{array}%
\right. $ \\ \hline
\end{tabular}%
$%
\end{tabular}%
\end{equation}%
Regarding the six 1-form gauge fields 
\begin{equation}
\begin{tabular}{llll}
$\mathcal{A}_{\mu }^{ai}=\left( \mathcal{A}_{\mu }^{a1},\mathcal{A}_{\mu
}^{a2}\right) $ & , & $\mathcal{A}_{\mu }^{a1}\equiv \mathcal{A}_{\mu }^{a},$
& $\mathcal{A}_{\mu }^{a2}\equiv \mathcal{K}_{\mu }^{a}$%
\end{tabular}%
\end{equation}%
three of them; say $\mathcal{A}_{\mu }^{a1}=\mathcal{G}_{\mu a}^{\left( 
{\scriptsize 11D}\right) }$, are Kaluza Klein type obtained from the metric
reduction; and the three $\mathcal{A}_{\mu }^{a2}$ others follow from the
reduction of $\mathcal{C}_{MNP}^{\left( {\scriptsize 11D}\right) }$ as, 
\begin{equation}
\mathcal{C}_{\mu \left[ ab\right] }^{\left( {\scriptsize 11D}\right)
}=\varepsilon _{abc}\mathcal{K}_{\mu }^{c}
\end{equation}%
where $\varepsilon _{abc}$ is the usual completely antisymmetric tensor of
the real \emph{3D} space. These fields are associated with the three gauge
particles given by the wrapping of M2 brane on the three 2-cycles of $T^{3}$
as illustrated on the following table,%
\begin{equation}
\begin{tabular}{l}
$%
\begin{tabular}{l|l|l|l|l|l|l|l|l||lll|}
\cline{2-12}
& \multicolumn{8}{|l}{$\ \ \ \ \ \mathcal{M}_{8}$ {\small space time}} & 
\multicolumn{3}{||l|}{\ \ \ $\ \ \ \ \ \ \ \ T^{3}$} \\ \cline{2-12}
& 0 & 1 & 2 & 3 & 4 & 5 & 6 & 7 & 8 & 9 & 10 \\ \hline
\multicolumn{1}{|l|}{M2} & $\times $ & $\circ $ & $\mathbf{\circ }$ & $%
\mathbf{\circ }$ & $\mathbf{\circ }$ & $\mathbf{\circ }$ & $\mathbf{\circ }$
& $\mathbf{\circ }$ & $\left\{ 
\begin{array}{c}
\mathbf{\circ } \\ 
\times \\ 
\times%
\end{array}%
\right. $ & $\left. 
\begin{array}{c}
\mathbf{\times } \\ 
\circ \\ 
\times%
\end{array}%
\right. $ & $\left. 
\begin{array}{c}
\mathbf{\times } \\ 
\times \\ 
\circ%
\end{array}%
\right. $ \\ \hline
\multicolumn{1}{|l|}{M5} & $\mathbf{\circ }$ & $\mathbf{\circ }$ & $\mathbf{%
\circ }$ & $\mathbf{\times }$ & $\times $ & $\times $ & $\times $ & $\times $
& $\left\{ 
\begin{array}{c}
\mathbf{\times } \\ 
\circ \\ 
\circ%
\end{array}%
\right. $ & $\left. 
\begin{array}{c}
\mathbf{\circ } \\ 
\times \\ 
\circ%
\end{array}%
\right. $ & $\left. 
\begin{array}{c}
\mathbf{\circ } \\ 
\circ \\ 
\times%
\end{array}%
\right. $ \\ \hline
\end{tabular}%
$%
\end{tabular}%
\end{equation}%
Before proceeding let us give some useful details.

\emph{8D }$\mathcal{N}=2$ \emph{supergravity fields}\newline
Under reduction to eight dimensions, the initial $128+128$ degrees of
freedom of \emph{11D} supergravity decomposes into various $SO\left(
1,7\right) $ representations. For the fermionic sector, we have 
\begin{equation}
128=\left( 2\times \left[ 6\times 8-8\right] \right) +\left( 6\times
8\right) .
\end{equation}%
The first block describes the degrees of freedom of two \emph{8D} Ravita
Schwinger fields $\psi _{\alpha \mu }^{A},$ $\bar{\psi}_{\dot{\alpha}\mu A}$%
\ ($A=1,2$) and the second one captures the degrees of freedom of $\left(
2\times 3\right) $ gauginos $\lambda _{\alpha }^{Aa},$ $\bar{\lambda}_{\dot{%
\alpha}}^{Aa}$. For the \emph{128} bosonic degrees of freedom; they
decompose as follows 
\begin{equation}
128=\left( 1+20\right) +20+\left( 3\times 15\right) +\left( 2\times 3\times
6\right) +\left( 5+1\right)
\end{equation}%
where $\left( 1+20\right) $ stand for the dilaton $\sigma $ and the metric
fields $G_{\mu \nu }$ and the second $20$ for the antisymmetric $\mathcal{C}%
_{\mu \nu \rho }$. The $3\times 15$ are the degrees of freedom of the $%
\mathcal{B}_{\mu \nu }^{a}$ triplet, the number $2\times 3\times 6$ describe
two triplets of gauge fields $\mathcal{A}_{\mu }^{ai}$ and finally $\left(
5+1\right) $ stand for the scalars $\varphi ^{\left( ab\right) },$ $%
\vartheta $. The later follow respectively from the reduction of the metric
and the 3-form gauge field on $T^{3}$ leading to the quintet $G_{ab}\equiv
\varphi ^{\left( ab\right) }$ and the singlet $\mathcal{C}_{abc}\sim 
\mathcal{\varepsilon }_{abc}\vartheta $. In summary, the bosonic content of 
\emph{8D }$\mathcal{N}=\left( 2,2\right) $ supergravity is, 
\begin{equation}
\begin{tabular}{lllll}
$\mathcal{G}_{\mu \nu },$ & $\mathcal{C}_{\mu \nu \rho },$ & $\mathcal{B}%
_{\mu \nu }^{a},$ & $\mathcal{A}_{\mu }^{ai},$ & $\varphi ^{\left( ab\right)
},$ $\ \sigma ,$ $\ \vartheta ,$%
\end{tabular}
\label{sm}
\end{equation}%
with $\sum_{a=1}^{3}\varphi ^{\left( aa\right) }=0$. Notice that in addition
to: (\textbf{1}) the seven scalar fields $\left\{ \phi \right\} $ that
parameterize the moduli space $\frac{SL\left( 3,R\right) }{SO\left( 3\right) 
}\times \frac{SL\left( 2,R\right) }{SO\left( 2\right) }$ $\sim $ $\frac{%
SU\left( 1,3\right) }{SU\left( 2\right) }\times \frac{SU\left( 1,1\right) }{%
U\left( 1\right) }$ to be discussed in details later on; and (\textbf{2})
the \emph{8D} graviton $\mathcal{G}_{\mu \nu }$ with scalar curvature $%
\mathcal{R}_{{\scriptsize 8}}$ and energy density,%
\begin{equation}
\begin{tabular}{l}
$\mathcal{L}_{{\scriptsize gravity}}=-\frac{1}{16\pi G_{8}}\int_{\mathcal{M}%
_{8}}d^{8}x\text{ }\sqrt{-\mathcal{G}}\text{ }\mathcal{R}_{{\scriptsize 8}}$%
\end{tabular}
\label{1}
\end{equation}%
with $\mathcal{G}=\det \mathcal{G}_{\mu \nu }$, we have moreover the
following:\newline
(\textbf{a}) the antisymmetric field $\mathcal{C}_{\mu \nu \rho }$ defining
a real gauge 3-form $\mathcal{C}_{{\scriptsize 3}}=\frac{1}{3!}dx^{\mu }%
{\small \wedge }dx^{\nu }{\small \wedge }dx^{\rho }\mathcal{C}_{\mu \nu \rho
}$ together with the 4-form field strength $\mathcal{F}_{{\scriptsize 4}}=d%
\mathcal{C}_{{\scriptsize 3}}$ and its magnetic dual $\mathcal{\tilde{F}}_{%
{\scriptsize 4}}=$ $^{\star }\mathcal{F}_{{\scriptsize 4}}$. The lagrangian
density describing the coupled dynamics of this field reads in general as
follows 
\begin{equation}
\begin{tabular}{ll}
$\mathcal{L}_{{\scriptsize 3}\text{{\scriptsize -}}{\scriptsize form}}=$ & $%
\frac{1}{32\pi G_{8}}\int_{\mathcal{M}_{8}}\sqrt{-\mathcal{G}}$ $\mathcal{N}%
_{FF}^{\left( {\small 2}\right) }\left( \phi \right) $ $\mathcal{F}_{\mu
_{1}...\mu _{4}}\mathcal{F}^{\mu _{1}...\mu _{4}}$%
\end{tabular}%
\end{equation}%
where the second term is topological. Implementing the duality relation $%
\mathcal{\tilde{F}}_{\mu _{5}...\mu _{8}}$ $\sim $ $\frac{1}{8!}\varepsilon
_{\mu _{1}...\mu _{8}}$ $\mathcal{F}^{\mu _{1}...\mu _{4}}$ by a Lagrange
multiplier $\mathcal{N}_{\tilde{F}\tilde{F}}^{\left( {\small 2}\right) }$,
we end with the gauge field action 
\begin{equation}
\begin{tabular}{l}
$\mathcal{L}_{{\scriptsize 3}\text{{\scriptsize -}}{\scriptsize form}}=\frac{%
1}{32\pi G_{8}}\int_{\mathcal{M}_{8}}\text{ }\sqrt{-\mathcal{G}}\text{ }%
\mathcal{N}_{ij}^{\left( {\small 2}\right) }\left( \phi \right) \text{ }%
\mathcal{F}_{\mu \nu \rho \sigma }^{i}\mathcal{F}^{\mu \nu \rho \sigma j}$%
\end{tabular}
\label{2}
\end{equation}%
where the field matrix $\mathcal{N}_{ij}^{\left( {\small 2}\right) }\left(
\phi \right) $, which can be factorized as $K_{i}^{m}\left( \phi \right)
\delta _{mn}K_{j}^{n}\left( \phi \right) $, provides the field coupling
metric for the kinetic terms. In this equation, we have also included the
topological term and set 
\begin{equation}
\begin{tabular}{llll}
$\mathcal{F}_{{\scriptsize 4}}^{i}=\left( 
\begin{array}{c}
\mathcal{F}_{{\scriptsize 4}} \\ 
\mathcal{\tilde{F}}_{{\scriptsize 4}}%
\end{array}%
\right) $ & $,$ & $\mathcal{F}_{{\scriptsize 4}i}=\mathcal{N}_{ij}^{\left( 
{\small 2}\right) }\mathcal{F}_{{\scriptsize 4}}^{j}$ & ,%
\end{tabular}%
\end{equation}%
transforming as an $SL\left( 2,R\right) $ doublet.\newline
(\textbf{b}) three antisymmetric gauge fields $\mathcal{B}_{\mu \nu }^{a}$
defining a triplet of real gauge 2-forms $\mathcal{B}_{2}^{a}$ with field
strengths $\mathcal{F}_{3}^{a}=d\mathcal{B}_{2}^{a}$ and magnetic duals $%
\mathcal{\tilde{F}}_{5|a}=$ $^{\star }\left( \mathcal{F}_{3}^{a}\right) $.
The lagrangian density describing their coupled dynamics reads as follows%
\begin{equation}
\begin{tabular}{ll}
$\mathcal{L}_{{\scriptsize 2}\text{{\scriptsize -}}{\scriptsize form}}=\frac{%
1}{32\pi G_{8}}\int_{\mathcal{M}_{8}}\sqrt{-\mathcal{G}}$ $\mathcal{N}%
_{ab}^{\left( {\small 1}\right) }\left( \phi \right) \mathcal{F}_{\mu \nu
\rho }^{a}\mathcal{F}^{\mu \nu \rho b}$ & ,%
\end{tabular}
\label{3}
\end{equation}%
with field metric $\mathcal{N}_{ab}^{\left( {\small 1}\right) }\left( \phi
\right) $ that can be factorized like $L_{a}^{c}\left( \phi \right) \delta
_{cd}L_{b}^{d}\left( \phi \right) $.\newline
(\textbf{c}) six real gauge fields $\mathcal{A}_{\mu }^{ai}$ defining six
real 1-forms $\mathcal{A}^{ai}=dx^{\mu }\mathcal{A}_{\mu }^{ai}$ with field
strengths $\mathcal{F}_{2}^{ai}=d\mathcal{A}^{ai}$ and magnetic duals $%
\mathcal{\tilde{F}}_{6}^{ai}=$ $^{\star }\left( \mathcal{F}_{2}^{ai}\right) $%
. For later physical interpretation, it is interesting to set, 
\begin{equation}
\begin{tabular}{llll}
$\mathcal{A}_{\mu }^{ai}=\left( 
\begin{array}{c}
\mathcal{A}_{\mu }^{a} \\ 
\mathcal{K}_{\mu }^{a}%
\end{array}%
\right) $ & , & $\mathcal{F}_{\mu \nu }^{ai}=\left( 
\begin{array}{c}
\mathcal{F}_{\mu \nu }^{a} \\ 
\mathcal{H}_{\mu \nu }^{a}%
\end{array}%
\right) $ & .%
\end{tabular}%
\end{equation}%
The lagrangian density of these gauge fields reads as follows%
\begin{equation}
\begin{tabular}{ll}
$\mathcal{L}_{{\scriptsize 1-form}}=\frac{1}{32\pi G_{8}}\int_{\mathcal{M}%
_{8}}\sqrt{-\mathcal{G}}\text{ }\mathcal{N}_{ai,bj}^{\left( {\small 0}%
\right) }\left( \phi \right) \text{ }\mathcal{F}_{\mu \nu }^{ai}\mathcal{F}%
^{\mu \nu bj}$ & .%
\end{tabular}
\label{4}
\end{equation}%
Notice that because of the factorization of the moduli space, the field
coupling $\mathcal{N}_{ai,bj}^{\left( {\small 0}\right) }$ decomposes as
well like $\mathcal{N}_{ab}^{\left( {\small 0}\right) }\times \mathcal{N}%
_{ij}^{\left( {\small 0}\right) }$. Notice also that the two internal
indices $\left( a,i\right) $ carried by the above gauge field strengths
refer to $SL\left( 3,R\right) \times SL\left( 2,R\right) $ representations.
We have,%
\begin{equation}
\begin{tabular}{lllllllllll}
\hline
{\small fields strenghts} & : & $\mathcal{F}_{4}^{i}$ & , & $\mathcal{F}%
_{3|a}$ & , & $\mathcal{\tilde{F}}_{5}^{a}$ & , & $\mathcal{F}_{2}^{ai}$ & ,
& $\mathcal{\tilde{F}}_{6|ai}$ \\ 
$SL\left( {\small 3}\right) \times SL\left( {\small 2}\right) $ & : & $%
\left( {\small 1,2}\right) $ & , & $\left( {\small 3}^{\prime }{\small ,1}%
\right) $ & , & $\left( {\small 3,1}\right) $ & , & $\left( {\small 3,2}%
\right) $ & , & $\left( {\small 3}^{\prime }{\small ,2}^{\prime }\right) $
\\ \hline
\end{tabular}%
\end{equation}%
For simplicity, we will sometimes refer collectively to these field
strengths as $\mathcal{F}_{p+2}^{I}=d\mathcal{A}_{p+1}^{I}$, $\mathcal{%
\tilde{F}}_{8-p-2|I}=$ $^{\star }\left( \mathcal{F}_{p+2}^{I}\right) $ with $%
\mathcal{A}_{p+1}^{I}$ the gauge $\left( p+1\right) $-forms taking values on 
$SL\left( 3\right) \times SL\left( 2\right) $ representations designated by
the $I$ index. These gauge invariant fields are associated with p-branes
(anti-p-branes) having electric charges $q_{I}$ and magnetic ones $p^{I}$
given by the generic relations 
\begin{equation}
\begin{tabular}{llll}
$p^{I}=\int_{\Sigma _{p+2}}\mathcal{F}_{p+2}^{I}$ & , & $q_{I}=\int_{\tilde{%
\Sigma}_{p+2}}\mathcal{\tilde{F}}_{6-p|I}$ & .%
\end{tabular}%
\end{equation}%
where the cycles $\Sigma _{p+2}$ and the dual $\tilde{\Sigma}_{6-p}$ may be
thought of as given by the spheres $S^{p+2}$ and $S^{6-p}$ respectively.

\emph{Brane configurations }\newline
Along with the M2 brane and the M2/S$^{1}$ as well as the M2/T$^{2}$ wrapped
geometries, we also have wrapped configurations induced by the M5 brane.
Since M5 is the magnetic dual of M2, the corresponding wrapped
configurations are dual to the ones associated with the membrane. In the
case of \emph{8D} $\mathcal{N}=\left( 2,2\right) $ supergravity, the
electric /magnetic duality that relates pairs of black p- and q- branes
requires $p+q=4$ from which we read the various black brane configurations
in \emph{8D}:\newline
(\textbf{i}) there are six black holes given by wrapping M2/T$^{2}$; these
black holes have magnetic charges $P^{ai}$ and transform in the
bi-fundamental of $SL\left( 3,R\right) \times SL\left( 2,R\right) $,%
\begin{equation}
\begin{tabular}{lllll}
$P^{ai}=\int_{S^{2}}\mathcal{F}_{2}^{ai}$ & $,$ & $\mathcal{N}%
_{aibj}^{\left( {\small 1}\right) }\mathcal{F}_{2}^{bi}=\mathcal{F}_{2|ai}$
& , & $\mathcal{N}_{aick}^{\left( {\small 1}\right) }$ $\mathcal{N}^{\left( 
{\small 1}\right) ckbj}=\delta _{a}^{b}\delta _{i}^{j}$%
\end{tabular}%
\end{equation}%
(\textbf{ii}) six black 4-branes obtained by wrapping M5/S$^{1}$; these
black objects are the dual of the black holes and carry the electric
charges, 
\begin{equation}
\begin{tabular}{ll}
$Q_{ai}=\int_{S^{6}}\mathcal{\tilde{F}}_{6|ai}$ & ,%
\end{tabular}%
\end{equation}%
with $P^{ai}Q_{bj}\sim n\delta _{b}^{a}\delta _{j}^{i}$ and where $n$ an
integer.\newline
(\textbf{iii}) three black strings obtained by wrapping M2/S$^{1}$; they are
magnetically charged,%
\begin{equation}
\begin{tabular}{ll}
$p_{a}=\int_{S^{3}}\mathcal{F}_{3|a}$ & ,%
\end{tabular}%
\end{equation}%
(\textbf{iv}) three black 3-branes following from the wrapping M5/T$^{2}$;
their electric charge reads as follows, 
\begin{equation}
\begin{tabular}{lll}
$q^{a}=\int_{S^{5}}\mathcal{\tilde{F}}_{5}^{a}$ & , & $q^{a}p_{b}\sim
n\delta _{b}^{a}.$%
\end{tabular}%
\end{equation}%
these are the dual of the black strings.\newline
(\textbf{v}) a dyonic black 2-brane given by the fundamental M2 and the
wrapped M5/T$^{3}$. Its electric $e$ and magnetic $g$ charges are as follows,%
\begin{equation}
\begin{tabular}{lllll}
$h^{i}=\int_{S^{4}}\mathcal{F}_{4}^{i}$ & , & $h^{i}=\left( g,e\right) $ & ,
& $eg\sim n.$%
\end{tabular}%
\end{equation}%
From this analysis, we learn that the full abelian gauge symmetry of the 
\emph{8D} $\mathcal{N}=2$ supergravity is given by 
\begin{equation}
\begin{tabular}{ll}
$U_{\text{{\scriptsize M2}}}\left( 1\right) \otimes U_{\text{{\scriptsize %
M2/S}}^{\text{{\scriptsize 1}}}}^{3}\left( 1\right) \otimes U_{\text{%
{\scriptsize M2/T}}^{\text{{\scriptsize 2}}}}^{3}\left( 1\right) \otimes U_{%
\text{{\scriptsize KK}}}^{3}\left( 1\right) $ & ,%
\end{tabular}
\label{ga}
\end{equation}%
where $U_{\text{{\scriptsize M2}}}\left( 1\right) $ stands for the gauge
group associated with the gauge 3-form, $U_{\text{{\scriptsize M2/S}}^{\text{%
{\scriptsize 1}}}}^{3}\left( 1\right) $ for strings and $U_{\text{%
{\scriptsize M2/T}}^{\text{{\scriptsize 2}}}}^{3}\left( 1\right) \times U_{%
\text{{\scriptsize KK}}}^{3}\left( 1\right) $ for the gauge particles.

\section{Attractor eqs of black p-branes}

In this section, we study an \emph{unconstrained} parametrization of the
moduli space $\frac{SL\left( 3,R\right) }{SO\left( 3\right) }\times $ $\frac{%
SL\left( 2,R\right) }{SO\left( 2\right) }$ of the \emph{8D} maximal
supergravity. This parametrization is based on using field matrices in $%
SL\left( 3,R\right) \times SL\left( 2,R\right) $ and gauging out the $%
SO\left( 3\right) \times SO\left( 2\right) $ isometries of the moduli space.
Then, we examine the total expression of the effective scalar potential $%
\mathcal{V}_{eff}$ of the black p-branes and derive the general expression
of the attractor equations associated with the various black p- branes
configurations living in \emph{8D}.

\subsection{Moduli space of \emph{8D} supergravity}

\subsubsection{Scalar fields}

In addition to the gauge fields and gauginos, the eight dimensional $%
\mathcal{N}=\left( 2,2\right) $ supergravity multiplet (\ref{sm}) has \emph{%
seven} real scalar fields $\left\{ \phi ^{1},...,\phi ^{7}\right\} $
parameterizing a non trivial scalar manifold. The first \emph{six} scalars,
to be denoted like $\left\{ \sigma ,\varphi _{\left( ab\right) }\right\} ,$ $%
Tr\left( \varphi \right) =0$, have a geometric interpretation in M theory
compactification on the 3-torus; the \emph{seventh}, denoted as $\vartheta $%
, has rather a stringy origin as the value of the gauge 3-form $\mathcal{C}_{%
{\scriptsize MNP}}^{\left( {\scriptsize 11D}\right) }$ on $T^{3}$. These
scalars capture special features on maximal supergravity in \emph{8D}; in
particular the two useful properties reported below.\newline
First, the seven scalars $\phi ^{1},...,\phi ^{7}$ organize into two
irreducible multiplets $\varphi _{\left( ab\right) }\oplus \xi _{\left(
ij\right) }$ with $5+2$ field components with the property $\sum_{a}\varphi
_{\left( aa\right) }=0,$ $\sum_{i}\xi _{\left( ii\right) }=0$. The fields $%
\varphi _{\left( ab\right) }$ are given by the following \emph{real symmetric%
} and traceless $3\times 3$ matrix, 
\begin{equation}
\begin{tabular}{lll}
$S_{{\scriptsize 0}}=\left( 
\begin{array}{ccc}
\phi ^{1} & \phi ^{3} & \phi ^{4} \\ 
\phi ^{3} & \phi ^{2} & \phi ^{5} \\ 
\phi ^{4} & \phi ^{5} & \phi ^{0}%
\end{array}%
\right) $ & , & $S_{{\scriptsize 0}}^{T}=S_{{\scriptsize 0}}$%
\end{tabular}
\label{f}
\end{equation}%
where we have set ${\small \phi }^{0}={\small -}\phi ^{1}{\small -}\phi ^{2}$
since $TrS_{{\scriptsize 0}}=0$. In group theoretic language \textrm{\cite%
{10A}}, this $S_{{\scriptsize 0}}$ matrix is associated with a particular
real group element 
\begin{equation}
M_{{\scriptsize 0}}=\exp S_{{\scriptsize 0}}
\end{equation}%
of the $SL\left( 3,R\right) $ group manifold, $M_{{\scriptsize 0}}^{\ast
}=M_{{\scriptsize 0}}$, $\det M_{{\scriptsize 0}}=1$; but moreover $M_{%
{\scriptsize 0}}^{T}=M_{{\scriptsize 0}}$ due to $S_{{\scriptsize 0}}^{T}=S_{%
{\scriptsize 0}}$. By using the general result that each generic $SL\left(
n,R\right) $ matrix $M$ can be usually decomposed as the product $U_{%
{\scriptsize 0}}\times M_{{\scriptsize 0}}\times U_{{\scriptsize 0}}^{T}$ of
an orthogonal $SO\left( n\right) $ matrix $U_{{\scriptsize 0}}$ and a
symmetric $M_{{\scriptsize 0}}$ one, it follows then that $M_{{\scriptsize 0}%
}$ is just a representative matrix of the coset $SL\left( 3,R\right)
/SO\left( 3\right) $; that is a representative element of the class 
\begin{equation}
\begin{tabular}{llllll}
$M\equiv U^{T}MU$ & , & $M\in SL\left( 3,R\right) $ & , & $U\in SO\left(
3\right) $ & .%
\end{tabular}
\label{re}
\end{equation}%
The same analysis holds for the other two real fields $\xi _{\left(
ij\right) }$; they organize into a real symmetric and traceless $2\times 2$
matrix of the form%
\begin{equation}
\begin{tabular}{llll}
$P_{{\scriptsize 0}}=\left( 
\begin{array}{cc}
\sigma & \vartheta \\ 
\vartheta & -\sigma%
\end{array}%
\right) $ & , & $Q_{{\scriptsize 0}}=\exp P_{{\scriptsize 0}}$ & ,%
\end{tabular}
\label{fi}
\end{equation}%
with $\phi ^{6}=\sigma $, $\phi ^{7}=\vartheta $. Here also, the real $%
2\times 2$ matrix $Q_{{\scriptsize 0}}$ is a representative matrix of the
class $Q\equiv V^{T}QV$ with $Q\in SL\left( 2,R\right) $ and $V\in SO\left(
2\right) $. Therefore, the seven scalar fields $\left\{ \phi ^{1},...,\phi
^{7}\right\} $ of the \emph{8D} maximal supergravity, organized as in eqs(%
\ref{f}-\ref{fi}), parameterize the real seven dimension non compact moduli
space 
\begin{equation}
\frac{SL\left( 3,R\right) }{SO\left( 3\right) }\times \frac{SL\left(
2,R\right) }{SO\left( 2\right) }\text{ \ }\sim \text{ \ }\frac{SU\left(
1,2\right) }{SU\left( 2\right) }\times \frac{SU\left( 1,1\right) }{U\left(
2\right) }.  \label{ms}
\end{equation}%
The second property, we want to comment is that the scalars $\left\{ \phi
^{1},...,\phi ^{7}\right\} $ generate $\phi $- dependent couplings among the
components of the supergravity multiplet (\ref{sm}). Some of these couplings
are given by the scalar functions $\mathcal{N}_{IJ}^{\left( p\right) }\left(
\phi \right) $ encountered previously (\ref{2}-\ref{3}). The other couplings
are given by self interactions as well as the coupling to the gravity field $%
\mathcal{G}_{\mu \nu }$ as shown on the lagrangian density 
\begin{equation}
\begin{tabular}{l}
$\mathcal{S}_{{\scriptsize Scalars}}=\frac{1}{32\pi G_{8}}\int d^{8}x\text{ }%
\sqrt{-\det \mathcal{G}}\text{ }\mathcal{G}^{\mu \nu }$ $g_{IJ}\left( \phi
\right) $ $\partial _{\mu }\phi ^{I}\partial _{\nu }\phi ^{J}$%
\end{tabular}%
\end{equation}%
where $g_{IJ}\left( \phi \right) $ is the metric of (\ref{ms}).\newline
To deal with the various couplings of these scalar fields as well as the
effective potential of the black branes $\mathcal{V}_{eff}\left( \phi
\right) $ to be considered later on, we shall develop a formalism based on
the typical relations (\ref{re}) and to which we refer to as the \emph{%
unconstrained method}. This formalism relies on working with two real matrix
fields; namely a $3\times 3$ matrix field $\left( L_{ab}\right) $ and a $%
2\times 2$ matrix $\left( K_{ij}\right) $ that are valued in the $SL\left(
3,R\right) $ $\times $ $SL\left( 2,R\right) $ Lie group,%
\begin{equation}
\begin{tabular}{llll}
$L_{ab}\in SL\left( 3,R\right) \ $ & , & $K_{ij}\in SL\left( 2,R\right) \ $
& 
\end{tabular}
\label{lkt}
\end{equation}%
and think about the $SO\left( 3\right) \times SO\left( 2\right) $ isometry
of the moduli space as an auxiliary gauge symmetry captured by \emph{%
auxiliary} gauge fields $\mathcal{A}_{\mu }^{{\scriptsize SO}_{{\scriptsize 3%
}}{\scriptsize \times SO}_{{\scriptsize 2}}}$. In this set up, physical
observables are expressed in terms of the $L$ and $K$ matrices; but are $%
SO\left( 3\right) \times SO\left( 2\right) $ invariant. Let us give some
useful details.

\subsubsection{More on unconstrained method}

Being group elements of $SL\left( 3,R\right) $ $\times $ $SL\left(
2,R\right) $ group manifold, the real matrices $L$ and $K$ satisfy the group
theoretical constraint eqs,%
\begin{equation}
\begin{tabular}{llllll}
$L_{ac}\tilde{L}^{cb}=\delta _{a}^{b}$ & , & $L^{-1}=\tilde{L}$ & , & $\det
L=1$ &  \\ 
$K_{il}\tilde{K}^{lj}=\delta _{i}^{j}$ & , & $K^{-1}=\tilde{K}$ & , & $\det
K=1$ & 
\end{tabular}%
\end{equation}%
fixing two real degrees of freedom among the real $9+4$. The other $\left(
3+1\right) $ undesired variables are fixed by requiring the following
identifications under the $SO\left( 3\right) \times SO\left( 2\right) $
symmetry of the moduli space%
\begin{equation}
\begin{tabular}{llll}
$K\equiv V^{T}KV$ & , & $V\in SO\left( 2\right) $ &  \\ 
$L\equiv U^{T}LU$ & , & $U\in SO\left( 3\right) $ & 
\end{tabular}
\label{23}
\end{equation}%
where $V=\exp \eta {\large \tau }$ with $\tau \equiv \tau ^{2}$ given by eq(%
\ref{t2}) and $U=\exp \zeta _{a}T^{a}$ are gauge transformations with
respective gauge parameters $\eta =\eta \left( \phi \right) $ and $\zeta
_{a}=\zeta _{a}\left( \phi \right) $. Notice that the three $T^{a}$'s are
given by the following antisymmetric $3\times 3$ matrices,%
\begin{equation}
\begin{tabular}{lll}
$\mathcal{T}_{1}=\left( 
\begin{array}{ccc}
0 & 1 & 0 \\ 
-1 & 0 & 0 \\ 
0 & 0 & 0%
\end{array}%
\right) ,$ & $\mathcal{T}_{2}=\left( 
\begin{array}{ccc}
0 & 0 & 1 \\ 
0 & 0 & 0 \\ 
-1 & 0 & 0%
\end{array}%
\right) ,$ & $\mathcal{T}_{3}=\left( 
\begin{array}{ccc}
0 & 0 & 0 \\ 
0 & 0 & 1 \\ 
0 & -1 & 0%
\end{array}%
\right) .$%
\end{tabular}
\label{3g}
\end{equation}%
We also require that the matrix gradients $\nabla _{\mu }^{so_{{\scriptsize 3%
}}}L$ and $\nabla _{\mu }^{so_{{\scriptsize 2}}}K$, with $\nabla _{\mu
}^{so_{{\scriptsize n}}}=\partial _{\mu }-\mathcal{A}_{\mu }^{so_{%
{\scriptsize n}}}$, are gauge covariant under these transformations so that
their space time kinetic energies $\frac{1}{2}\sqrt{-\mathcal{G}}Tr\left( 
\mathcal{G}^{\mu \nu }\left( \nabla _{\mu }L\right) L^{-1}\left( \nabla
_{\nu }L\right) L^{-1}\right) $ $+$ $\frac{1}{2}\sqrt{-\mathcal{G}}Tr\left( 
\mathcal{G}^{\mu \nu }\left( \nabla _{\mu }K\right) K^{-1}\left( \nabla
_{\nu }K\right) K^{-1}\right) $ are gauge invariant. More precisely, we have%
\begin{equation}
\begin{tabular}{lll}
$\left( \nabla _{\mu }L\right) L^{-1}$ & $\equiv U^{T}\left[ \left( \nabla
_{\mu }^{so_{{\scriptsize 3}}}L\right) L^{-1}\right] U$ & , \\ 
$\left( \nabla _{\mu }K\right) K^{-1}$ & $\equiv V^{T}\left[ \left( \nabla
_{\mu }^{so_{{\scriptsize 2}}}K\right) K^{-1}\right] V$ & ,%
\end{tabular}
\label{24}
\end{equation}%
where the \emph{8D} vector fields $\left( \mathcal{A}_{\mu }^{so_{%
{\scriptsize 3}}},\mathcal{A}_{\mu }^{so_{{\scriptsize 2}}}\right) $ are
gauge fields associated with the $SO\left( 3\right) \times SO\left( 2\right) 
$ isometry of the moduli space. Under $SO\left( 3\right) \times SO\left(
2\right) $ change generated by the $\left( U,V\right) $ orthogonal matrices,
these gauge fields transform respectively as $\mathcal{A}_{\mu }^{so_{%
{\scriptsize 3}}}+U\partial _{\mu }U^{T}$, $\mathcal{A}_{\mu }^{so_{%
{\scriptsize 2}}}+V\partial _{\mu }V^{T}$. Notice also that the gauge fields 
$\mathcal{A}_{\mu }^{so_{{\scriptsize 3}}}$ and $\mathcal{A}_{\mu }^{so_{%
{\scriptsize 2}}}$ are auxiliary fields in the sense that they do not have
kinetic terms; the elimination of these fields through their equations of
motion allows to express them as functions of the L and K matrices and their
space time derivatives,%
\begin{equation}
\begin{tabular}{llll}
$\mathcal{A}_{\mu }^{so_{{\scriptsize 3}}}=F\left( L,\partial _{\mu
}L\right) $ & , & $\mathcal{A}_{\mu }^{so_{{\scriptsize 2}}}=F\left(
K,\partial _{\mu }K\right) $ & ,%
\end{tabular}%
\end{equation}%
which, up on substitution, induce non trivial self interactions amongst the
matrix fields leading to the metric of the moduli space $\frac{SL\left(
3,R\right) \times SL\left( 2,R\right) }{SO\left( 3\right) \times SO\left(
2\right) }$. \newline
The $SO\left( 3\right) \times SO\left( 2\right) $ identifications (\ref{23}-%
\ref{24}) can be explicitly illustrated by expressing the field matrices L
and K as Lie group elements like,%
\begin{equation}
\begin{tabular}{llll}
$L=\exp \varphi $ & , & $\ K=\exp \xi $ & $,$ \\ 
$U=\exp \zeta $ & , & $\ V=\exp \eta .{\large \tau }^{2}$ & $,$%
\end{tabular}
\label{par}
\end{equation}%
with $\varphi =\sum_{a,b}\left( \varphi _{ab}-\frac{1}{3}\chi \delta
_{ab}\right) T^{ab},$ $\xi =\sum_{m,n}\left( \xi _{mn}-\frac{1}{2}\varkappa
\delta _{MN}\right) {\large \tau }^{mn}$ which read also like, 
\begin{equation}
\begin{tabular}{llll}
$\varphi =\sum\limits_{A=1}^{8}\varphi _{A}\mathcal{T}^{A},$ & $\xi
=\sum\limits_{\alpha =1}^{3}\xi _{\alpha }\mathcal{\tau }^{\alpha },$ & $%
\zeta =\sum \zeta _{a}T^{a}$ & .%
\end{tabular}
\label{ant}
\end{equation}%
Here the eight traceless 3$\times $3 matrices $T^{ab}$ (or equivalently $%
\mathcal{T}^{A}$) are the generators of $SL\left( 3,R\right) $; they may be
split as $\left( 3+5\right) $ describing respectively $T^{\left[ ab\right]
}=\varepsilon ^{abc}T^{c}$ generating the subgroup $SO\left( 3\right) $ and $%
T^{\left( ab\right) }$ generating the $SL\left( 3,R\right) /SO\left(
3\right) $ manifold. The three $2\times 2$ traceless matrices ${\large \tau }%
^{mn}$ (or equivalently $\mathcal{\tau }^{\alpha }$) are the generators of $%
SL\left( 2,R\right) $; they split as $\left( 1+2\right) $ describing
respectively ${\large \tau }^{\left[ mn\right] }=\varepsilon ^{mn}\tau ^{2}$
generating $SO\left( 2\right) $ and ${\large \tau }^{\left( mn\right) }$
generating the space $SL\left( 2,R\right) /SO\left( 2\right) $. These
generators read as follows: 
\begin{equation}
\begin{tabular}{lll}
$\mathcal{\tau }^{1}=\left( 
\begin{array}{cc}
0 & 1 \\ 
1 & 0%
\end{array}%
\right) ,$ & $\mathcal{\tau }^{2}=\left( 
\begin{array}{cc}
0 & 1 \\ 
-1 & 0%
\end{array}%
\right) ,$ & $\mathcal{\tau }^{3}=\left( 
\begin{array}{cc}
1 & 0 \\ 
0 & -1%
\end{array}%
\right) $%
\end{tabular}
\label{t2}
\end{equation}%
with the relations%
\begin{equation}
\begin{tabular}{lll}
$\mathcal{\tau }^{1}={\large \tau }^{12}+{\large \tau }^{21}$ & , & $%
\mathcal{\tau }^{3}={\large \tau }^{11}-{\large \tau }^{22}$ \\ 
$\mathcal{\tau }^{2}={\large \tau }^{12}-{\large \tau }^{21}$ & , & $%
\mathcal{\delta }^{0}={\large \tau }^{11}+{\large \tau }^{22}$%
\end{tabular}%
\end{equation}%
where sometimes we also use the notation $\mathcal{\tau }^{3}=\mathcal{\tau }%
^{0}$. The net field variables parameterizing the typical coset manifolds $%
SL\left( n,R\right) /SO\left( n\right) $ may be obtained by decomposing the
adjoint representation of $SL\left( n,R\right) $ with respect to the
irreducible representations of $SO\left( n\right) $. We have%
\begin{equation}
\begin{tabular}{l}
$n^{2}-1=\frac{n\left( n-1\right) }{2}\oplus \frac{n\left( n+1\right) -2}{2}$%
\end{tabular}%
\end{equation}%
where $\frac{n\left( n-1\right) }{2}$ stands for $ad_{SO\left( n\right) }$
and $\frac{n\left( n+1\right) -2}{2}$ for the traceless symmetric
representation. Gauge symmetry under $SO\left( n\right) $ may be used to fix
the antisymmetric part $\sum_{a,b}\varphi _{\left[ ab\right] }T^{\left[ ab%
\right] }$ in the typical expansions (\ref{ant}) leaving free the real $%
\frac{n\left( n+1\right) -2}{2}$ variables. \newline
In $\mathcal{N}=\left( 2,2\right) $ supergravity where the role of $SL\left(
n,R\right) $ is played by the direct product $SL\left( 3,R\right) $ $\times $
$SL\left( 2,R\right) $, the decomposition with respect to $SO\left( 3\right)
\times SO\left( 2\right) $ reads as,%
\begin{equation}
\begin{tabular}{llll|llll}
\hline
$SL\left( 3,R\right) $ & $\supset $ & $SO\left( 3\right) $ &  &  & $SL\left(
2,R\right) $ & $\supset $ & $SO\left( 2\right) $ \\ 
$\ \ \ \ \ \ \ 8$ & $=$ & $3\oplus 5$ &  &  & $\ \ \ \ \ \ \ 3$ & $=$ & $%
1\oplus 2$ \\ \hline
\end{tabular}%
\end{equation}%
The $\left( 5+2\right) $ physical degrees of freedom $\varphi ^{\left(
ab\right) }$, $\xi ^{\left( ij\right) }$ parameterizing (\ref{ms}) may be
explicitly exhibited by solving the above $SO\left( 3\right) \times SO\left(
2\right) $ identifications (\ref{23}) to end with the gauge fixed
representatives $L_{{\scriptsize 0}}$ and $K_{{\scriptsize 0}}$ given by,%
\begin{equation}
\begin{tabular}{lll}
$L_{{\scriptsize 0}}=e^{\varphi _{_{_{0}}}},$ & $\varphi
_{_{0}}=\sum_{a,b}\varphi _{\left( ab\right) }T^{\left( ab\right) },$ & , \\ 
$K_{{\scriptsize 0}}=e^{\xi _{_{0}}},$ & $\xi _{_{0}}=\sum_{m,n}\xi _{\left(
mn\right) }{\large \tau }^{\left( mn\right) }$ & ,%
\end{tabular}
\label{lo}
\end{equation}%
where $\varphi _{\left( ab\right) }=\varphi _{\left( ba\right) }$, $\xi
_{\left( mn\right) }=\xi _{\left( nm\right) }$ and $\sum_{a}\varphi _{\left(
aa\right) }=0,$ $\sum_{m}\xi _{\left( mm\right) }=0$. For later use, we
rewrite the matrix $K_{{\scriptsize 0}}$ like,%
\begin{equation}
\begin{tabular}{lll}
$K_{{\scriptsize 0}}$ & $=\exp \left( \vartheta {\large \tau }^{_{1}}+\sigma 
{\large \tau }^{_{3}}\right) $ & .%
\end{tabular}
\label{ko}
\end{equation}%
These gauge fixed matrices should be compared with (\ref{f}-\ref{fi}).

\subsubsection{Maurer Cartan forms}

In the unconstrained parametrization of the moduli space (\ref{ms}), the
basic field variables are the matrices L and K obeying (\ref{23}-\ref{24}).
The variations of these field matrices are captured by the Maurer Cartan
1-forms $\Omega ^{{\scriptsize SL}_{{\scriptsize 3}}}\equiv \Omega $ and $%
\Omega ^{{\scriptsize SL}_{{\scriptsize 2}}}\equiv \omega $ \textrm{\cite%
{8B,10A}} living on the $SL\left( 3,R\right) \times SL\left( 2,R\right) $
group manifold. These real 1-forms,%
\begin{equation}
\begin{tabular}{ll}
$\Omega =-L\left( dL^{-1}\right) $ & , \\ 
$\omega =-K\left( dK^{-1}\right) $ & ,%
\end{tabular}
\label{m}
\end{equation}%
depend implicitly of the group parameters $\varphi ^{A},$ $\xi ^{\alpha }$
and their $d\varphi ^{A},$ $d\xi ^{\alpha }$ differentials and exhibit two
basic kinds of expansions; the first one with respect to the field
differential basis $\left\{ d\varphi ^{A},d\xi ^{\alpha }\right\} $ and the
second with respect to the generators $\left\{ \mathcal{T}_{A},\mathcal{\tau 
}_{\alpha }\right\} $ of the $sl\left( 3,R\right) \oplus sl\left( 2,R\right) 
$ Lie algebra.

\emph{Differential basis} $\left\{ d\varphi ^{A},d\xi ^{\alpha }\right\} $%
\newline
Substituting the matrices L and K by their respective expressions $\exp
\left( \sum_{A}\varphi _{A}T^{A}\right) $ and $\exp \left( \sum_{\alpha }\xi
_{\alpha }\tau ^{\alpha }\right) $ into eq(\ref{m}), we see that we can
expand these Maurer Cartan forms in a series as follows%
\begin{equation}
\begin{tabular}{llll}
$\Omega =\sum\limits_{A=1}^{8}d\varphi ^{A}\Omega _{A}$ & , & $\omega
=\sum\limits_{\alpha =1}^{3}d\xi ^{\alpha }\omega _{\alpha }$ & ,%
\end{tabular}
\label{mc}
\end{equation}%
with sections $\Omega _{A}=-L\mathcal{T}_{A}L^{-1}$ and $\omega _{\alpha
}=-K\tau _{\alpha }K^{-1}$ which are noting but the Lie group adjoint
actions on the $sl\left( 3,R\right) \oplus sl\left( 2,R\right) $ generators, 
\begin{equation}
\begin{tabular}{llll}
$\Omega _{A}=-e^{ad_{\varphi }}\mathcal{T}_{A}$ & , & $\omega _{\alpha
}=-e^{ad_{\xi }}\mathcal{\tau }_{\alpha }$ & .%
\end{tabular}%
\end{equation}%
These sections are real matrices that are respectively valued in the $%
sl\left( 3,R\right) $ and $sl\left( 2,R\right) $ Lie algebras as shown by
the traces $Tr\left( \Omega _{A}\right) =0$, $Tr\left( \omega _{\alpha
}\right) =0$. Notice also that thinking about $sl\left( 3,R\right) \oplus
sl\left( 2,R\right) $ as a vector space, the Maurer Cartan fields $\Omega $
and $\omega $ may be split as follows 
\begin{equation}
\begin{tabular}{ll}
$\Omega =\Omega ^{{\scriptsize SO}_{{\scriptsize 3}}}+\Omega ^{{\scriptsize %
SL}_{{\scriptsize 3}}{\scriptsize /SO}_{{\scriptsize 3}}}$ & , \\ 
$\omega =\omega ^{{\scriptsize SO}_{{\scriptsize 2}}}+\omega ^{{\scriptsize %
SL}_{{\scriptsize 2}}{\scriptsize /SO}_{{\scriptsize 2}}}$ & ,%
\end{tabular}%
\end{equation}%
where the terms $\Omega ^{{\scriptsize SO}_{{\scriptsize n}}}$ and $\Omega ^{%
{\scriptsize SL}_{{\scriptsize n}}{\scriptsize /SO}_{{\scriptsize n}}}$ are
respectively the Maurer Cartan forms associated with $SO\left( n\right) $
group and the coset space $SL\left( n,R\right) {\scriptsize /}SO\left(
n\right) $.

\emph{Lie algebra basis }$\left\{ \mathcal{T}_{A},\mathcal{\tau }_{\alpha
}\right\} $\emph{\ }\newline
Using specific properties of $SL\left( n,R\right) $ matrices; in particular
the adjoint action $e^{A}Be^{-A}=e^{ad_{A}}B$ with \ $ad_{A}B=AB-BA$ and
applying this to the field matrices $L=e^{\varphi },$ $K=e^{\xi }$, we can
express the sections $\Omega _{A}=-e^{ad_{\varphi }}T_{A}$ and $\omega
_{\alpha }=-e^{ad_{\xi }}\tau _{\alpha }$ as an infinite series like,%
\begin{equation}
\begin{tabular}{ll}
$\Omega _{A}=-\mathcal{T}_{A}-\sum\limits_{n=1}^{\infty }\frac{1}{n!}\left[
\varphi ,\left[ \varphi ,...\left[ \varphi ,\mathcal{T}_{A}\right] ...\right]
\right] _{n}$ & , \\ 
$\omega _{\alpha }=-\mathcal{\tau }_{\alpha }-\sum\limits_{n=1}^{\infty }%
\frac{1}{n!}\left[ \xi ,\left[ \xi ,...\left[ \xi ,\tau _{\alpha }\right] ...%
\right] \right] _{n}$ & .%
\end{tabular}%
\end{equation}%
Now substituting $\varphi =\sum_{B}\varphi _{B}\mathcal{T}^{B}$ and $\xi
=\sum_{\beta }\xi _{\beta }\tau ^{\beta }$ in these relations and using $%
\left[ \varphi ,\mathcal{T}_{A}\right] =F_{BA}^{C}\varphi ^{B}\mathcal{T}%
_{C} $ and $\left[ \xi ,\mathcal{\tau }_{\alpha }\right] =\mathrm{f}_{\beta
\alpha }^{\gamma }\varphi ^{\beta }\mathcal{\tau }_{\gamma }$ with $%
F_{BA}^{C}$ and $\mathrm{f}_{\beta \alpha }^{\gamma }$ standing for the
structure constants of the $sl\left( 3,R\right) $ and $sl\left( 2,R\right) $
Lie algebras, we learn that the $\Omega _{A}$ and $\omega _{\alpha }$
matrices may be expanded in terms of the generators $\left\{ \mathcal{T}%
^{B};\tau ^{\beta }\right\} $ as follows 
\begin{equation}
\begin{tabular}{llll}
$\Omega _{A}=\sum\limits_{B}\Theta _{A}^{B}\mathcal{T}_{B}$ & , & $\omega
_{\alpha }=\sum_{\beta }\theta _{\alpha }^{\beta }\mathcal{\tau }_{\beta }$
& .%
\end{tabular}%
\end{equation}%
Seen that these expansions are useful in dealing with attractor eqs of black
p-branes; let us collect here below the relevant relations:\newline
(\textbf{i}) the Maurer Cartan 1-forms $\Omega $ and $\omega $ (\ref{mc})
may be expanded into different ways; either with respect to the differential
form basis as%
\begin{equation}
\begin{tabular}{llll}
$\Omega =\sum\limits_{A=1}^{8}d\varphi ^{A}\Omega _{A}$ & , & $\omega
=\sum\limits_{\alpha =1}^{3}d\xi ^{\alpha }\omega _{\alpha }$ & ,%
\end{tabular}
\label{29}
\end{equation}%
or with respect to the Lie algebra generators like 
\begin{equation}
\begin{tabular}{llll}
$\Omega =\sum\limits_{B=1}^{8}\Delta ^{B}\mathcal{T}_{B}$ & , & $\omega
=\sum\limits_{\beta =1}^{3}\lambda ^{\beta }\mathcal{\tau }_{\beta }$ & ,%
\end{tabular}%
\end{equation}%
In the first expansion $\Omega _{A}$ and $\omega _{\alpha }$ are matrices
valued in the Lie algebras and in the second development $\Delta ^{B}$ and $%
\lambda ^{\beta }$ are real 1-forms. Combining the two expansions, we get, 
\begin{equation}
\begin{tabular}{llll}
$\Omega =\sum\limits_{A,B=1}^{8}d\varphi ^{A}\Theta _{A}^{B}\mathcal{T}_{B}$
& , & $\omega =\sum\limits_{\alpha ,\beta =1}^{3}d\xi ^{\alpha }\theta
_{\alpha }^{\beta }\mathcal{\tau }_{\beta }$ & ,%
\end{tabular}%
\end{equation}%
with%
\begin{equation}
\begin{tabular}{llll}
$\Delta ^{B}=\sum_{A}d\varphi ^{A}\Theta _{A}^{B}$ & , & $\lambda ^{\beta
}=\sum d\xi ^{\alpha }\theta _{\alpha }^{\beta }$ &  \\ 
$\Omega _{A}=\sum \Theta _{A}^{B}\mathcal{T}_{B}$ & , & $\omega _{\alpha
}=\sum \theta _{\alpha }^{\beta }\mathcal{\tau }_{\beta }$ & 
\end{tabular}
\label{fo}
\end{equation}%
and%
\begin{equation}
\begin{tabular}{llll}
$\Theta _{A}^{B}=-\left( e^{ad\varphi }\right) _{A}^{B}$ & , & $\theta
_{\alpha }^{\beta }=-\left( e^{ad\xi }\right) _{\alpha }^{\beta }$ & .%
\end{tabular}%
\end{equation}%
(\textbf{ii}) In the Cartan Weyl basis $\left\{ H_{i},E^{\pm \eta }\right\}
\oplus \left\{ \tau ^{0},\tau ^{\pm }\right\} $ of $sl\left( 3,R\right)
\oplus sl\left( 2,R\right) $, the Maurer Cartan fields $\Omega $ and $\omega 
$ read as 
\begin{equation}
\begin{tabular}{ll}
$\Omega =\sum\limits_{i}\Delta ^{i}H_{i}+\sum\limits_{\eta }\left( \Delta
^{-\eta }E^{+\eta }+\Lambda ^{+\eta }E^{-\eta }\right) $ & , \\ 
$\omega =\sum \lambda ^{0}\tau ^{0}+\sum \lambda ^{-}\tau ^{+}+\sum \lambda
^{+}\tau ^{-}$ & .%
\end{tabular}%
\end{equation}%
where $\eta $ refers to the positive roots of $sl\left( 3,R\right) $ and
where $\left( \Delta ^{i},\Delta ^{\pm \eta }\right) $ and $\left( \lambda
^{0},\lambda ^{\pm }\right) $ are differential forms given by (\ref{fo}).%
\newline
(iii) To solve the attractor eqs, we will use different representations to
deal with the Maurer Cartan forms; in particular the above ones but also $%
\Omega =\sum d\varphi ^{ab}\Omega _{ab}$, $\omega =\sum d\xi ^{mn}\omega
_{mn}$ where $\Omega _{ab}$ and $\omega _{mn}$ are related to $\Omega _{A}$
and $\omega _{\alpha }$\ as $\Omega _{ab}=\sum \Omega _{A}\mathcal{T}%
_{ab}^{A}$ and $\omega _{mn}=\sum \omega _{\alpha }\mathcal{\tau }%
_{mn}^{\alpha }$. Similarly, we also have $d\varphi ^{A}=\sum d\varphi ^{ab}%
\mathcal{T}_{ab}^{A}$\ and $d\xi ^{\alpha }=\sum d\xi ^{mn}\mathcal{\tau }%
_{mn}^{\alpha }$. In this basis, the Maurer Cartan forms associated with the 
$SO\left( n\right) $\ and $SL\left( n,R\right) /SO\left( n\right) $ are
given by the antisymmetric and symmetric parts as shown below 
\begin{equation}
\begin{tabular}{ll}
$\Omega =\sum\limits_{a,b}d\varphi ^{\left[ ab\right] }\Omega _{\left[ ab%
\right] }^{{\scriptsize SO}_{{\scriptsize 3}}}+\sum\limits_{a,b}d\varphi
^{\left( ab\right) }\Omega _{\left( ab\right) }^{{\scriptsize SL}_{%
{\scriptsize 3}}{\scriptsize /SO}_{{\scriptsize 3}}}$ & , \\ 
$\omega =\sum\limits_{m,n}d\xi ^{\left[ mn\right] }\omega _{\left[ mn\right]
}^{{\scriptsize SO}_{{\scriptsize 2}}}+\sum\limits_{m,n}d\xi ^{\left(
mn\right) }\omega _{\left( mn\right) }^{{\scriptsize SL}_{{\scriptsize 2}}%
{\scriptsize /SO}_{{\scriptsize 2}}}$ & .%
\end{tabular}%
\end{equation}%
In this representation, the $SO\left( 3\right) \times SO\left( 2\right) $
symmetry of the moduli space may be gauged out by setting $\Omega _{\left[ ab%
\right] }^{{\scriptsize SO}_{{\scriptsize 3}}}=0,$ $\omega _{\left[ mn\right]
}^{{\scriptsize SO}_{{\scriptsize 2}}}=0$ leaving only the desired
components $\Omega _{\left( ab\right) }^{{\scriptsize SL}_{{\scriptsize 3}}%
{\scriptsize /SO}_{{\scriptsize 3}}}$ and $\omega _{\left( mn\right) }^{%
{\scriptsize SL}_{{\scriptsize 2}}{\scriptsize /SO}_{{\scriptsize 2}}}$.

\subsection{Attractor equations}

Attractor equations of black objects in \emph{8D} maximal supergravity are
obtained by minimizing their effective potential $\mathcal{V}_{eff}\left(
\phi \right) $; its Hessian matrix is then positive \textrm{\cite{8A}}. This
scalar potential depend on the coordinates $\left\{ \phi ^{I}\right\} $ of
the moduli space of the theory. So, attractor eqs follow from $\delta
V_{eff}=\sum \frac{\partial V}{\partial \phi ^{I}}\delta \phi ^{I}$. For
arbitrary variations $\delta \phi ^{I}$, we have the following constraint
relations: 
\begin{equation}
\begin{tabular}{llllll}
$\frac{\partial \mathcal{V}_{eff}}{\partial \phi ^{I}}=0$ & , & $\det \left( 
\frac{\partial ^{2}\mathcal{V}_{eff}}{\partial \phi ^{I}\partial \phi ^{J}}%
\right) >0$ & , & $I,J=1,...,N$ & ,%
\end{tabular}
\label{tr}
\end{equation}%
whose solutions fix the values of the field moduli at the black object near
horizon geometry in terms of the charges $q$ and $p$; i.e $\phi _{I}=\phi
_{I}\left( p,q\right) $. \newline

\subsubsection{Effective potential}

In \emph{8D}\ maximal supergravity where lives several kinds of black
p-branes, the total effective potential, induced from the kinetic energies
of the gauge fields strengths at the horizon, is given by the sum over
individual components $\mathcal{V}_{p}$ associated with each black p-brane
as given below,%
\begin{equation}
\mathcal{V}_{eff}=\left( \mathcal{V}_{0}+\mathcal{V}_{4}\right) +\left( 
\mathcal{V}_{1}+\mathcal{V}_{3}\right) +\mathcal{V}_{2}\text{ }.  \label{pot}
\end{equation}%
The scalar components $\mathcal{V}_{p}$, which are related by the
electric/magnetic duality property $\mathcal{\tilde{V}}_{p}=\mathcal{V}%
_{4-p} $, are functions of the scalar fields (\ref{f}) and the charges $%
\left\{ g^{I},e_{I}\right\} $ of the branes, 
\begin{equation}
\mathcal{V}_{p}=\mathcal{V}_{p}\left( \phi ^{1},...,\phi
^{7};g^{I},e_{I}\right) .
\end{equation}%
In the unconstrained formulation of the moduli space, the effective
potential dependence in the $\phi $'s is realized through the field matrices 
$L_{ab}=L_{ab}\left( \phi \right) ,$ $K_{ij}=K_{ij}\left( \phi \right) $ so
that the $\mathcal{V}_{p}$ components are functionals like, 
\begin{equation}
\begin{tabular}{ll}
$\mathcal{V}_{p}=\mathcal{V}_{p}\left[ L\left( \phi \right) ,K\left( \phi
\right) ;g^{I},e_{I}\right] $ & ,%
\end{tabular}%
\end{equation}%
with the symmetry property%
\begin{equation}
\begin{tabular}{ll}
$\mathcal{V}_{p}\left[ L,K\right] =\mathcal{V}_{p}\left[ L^{\prime
},K^{\prime }\right] $ & ,%
\end{tabular}%
\end{equation}%
where%
\begin{equation}
\begin{tabular}{ll}
$L^{\prime }=U^{T}LU,$ & $K^{\prime }=V^{T}KV,$%
\end{tabular}%
\end{equation}%
are gauge transformations expressing invariance under $SO\left( 3\right)
\times SO\left( 2\right) $ isometry of the moduli space. The individual
potentials $\mathcal{V}_{0},$ $\mathcal{V}_{1}$ and $\mathcal{\tilde{V}}_{0}=%
\mathcal{V}_{4},$ $\mathcal{\tilde{V}}_{1}=\mathcal{V}_{3}$ are explicitly
expressed like 
\begin{equation}
\begin{tabular}{llll}
$\mathcal{V}_{0}=\frac{1}{2}\sum X^{ai}\delta _{ab}\delta _{ij}X^{bj}$ & , & 
$\mathcal{\tilde{V}}_{0}=\frac{1}{2}\sum \tilde{X}_{ai}\delta ^{ab}\delta
^{ij}\tilde{X}_{bj}$ & , \\ 
$\mathcal{V}_{1}=\frac{1}{2}\sum Y^{a}\delta _{ab}Y^{b}$ & , & $\mathcal{%
\tilde{V}}_{1}=\frac{1}{2}\sum \tilde{Y}_{a}\delta ^{ab}\tilde{Y}_{b}%
\mathcal{\ }$ & .%
\end{tabular}
\label{po}
\end{equation}%
In these relations, $X^{ai}$ and $\tilde{X}_{ai}$ are respectively the
dressed central charges of the black holes and their 4-branes dual while the
fields $Y^{a}$, $\tilde{Y}_{a}$ are the dressed central charges of the black
strings and their 3-branes dual. We also have%
\begin{equation}
\begin{tabular}{ll}
$\mathcal{V}_{2}=\frac{1}{2}Z_{el}^{2}+\frac{1}{2}Z_{mag}^{2}$ & ,%
\end{tabular}
\label{pt}
\end{equation}%
describing the effective potential of the black membrane. To exhibit the $%
SO\left( 3\right) \times SO\left( 2\right) $ symmetry of this potential as
in (\ref{po}), it is interesting to think about $\mathcal{V}_{2}$ as given
by the following symplectic form%
\begin{equation}
\begin{tabular}{ll}
$\mathcal{V}_{2}=\sum Z^{i}\delta _{ij}Z^{j}$ & $,$%
\end{tabular}%
\end{equation}%
with $Z^{i}=\left( Z_{mag},Z_{el}\right) $. Moreover, by using the field
matrices $L$, $K$ and the bare electric and magnetic charges associated with
the various fields strengths of the supergravity theory, we can express the
above dressed charges as follows%
\begin{equation}
\begin{tabular}{llll}
$X^{ai}=\sum P^{bj}$ $L_{b}^{a}K_{j}^{i}$ & $,$ & $\tilde{X}_{ai}=\sum
\left( L^{-1}\right) _{a}^{b}\left( K^{-1}\right) _{i}^{j}$ $Q_{bj}$ & $,$%
\end{tabular}
\label{drs}
\end{equation}%
and%
\begin{equation}
\begin{tabular}{llll}
$Y^{a}=\sum p^{b}$ $L_{b}^{a}$ & , & $\tilde{Y}_{a}=\sum \left(
L^{-1}\right) _{a}^{b}q_{b}$ & ,%
\end{tabular}
\label{ya}
\end{equation}%
as well as 
\begin{equation}
\begin{tabular}{ll}
$Z^{i}=\sum K_{j}^{i}$ $h^{j}$ & .%
\end{tabular}
\label{zc}
\end{equation}%
These dressed charges obey the typical $SO\left( 3\right) \times SO\left(
2\right) $ symmetry properties 
\begin{equation}
\begin{tabular}{lllll}
$X^{ai}\equiv \sum X^{bj}U_{b}^{a}V_{j}^{i}$ & , & $Y^{a}\equiv \sum
Y^{b}U_{b}^{a}$ & , & $Z^{i}\equiv \sum Z^{j}V_{j}^{i}$%
\end{tabular}%
\end{equation}%
that are induced by the symmetric features satisfied by the fields matrices
L and K.

\subsubsection{Attractor equations}

To get the attractor equations of the black p-branes, we extremize the above
effective potential $\mathcal{V}_{eff}$ with respect to the scalar fields $%
\left\{ \phi ^{I}\right\} $. Since $\mathcal{V}_{eff}$ \ is a functional of
these scalar fields that can be either thought of as 
\begin{equation}
\mathcal{V}_{eff}=\mathcal{V}_{eff}\left[ L_{ab}\left( \phi \right)
,K_{ij}\left( \phi \right) \right]  \label{la}
\end{equation}%
or in terms of the dressed central charges given by eqs(\ref{drs}-\ref{zc}),%
\begin{equation}
\begin{tabular}{ll}
$\mathcal{V}_{eff}=\mathcal{V}_{eff}\left[ X^{ai}\left( \phi \right)
,Y^{a}\left( \phi \right) ,Z^{i}\left( \phi \right) ,\tilde{X}_{ai}\left(
\phi \right) ,\tilde{Y}_{a}\left( \phi \right) \right] $ & ,%
\end{tabular}
\label{al}
\end{equation}%
we can state its extremum in two different, but equivalent, ways. Below, we
shall refer to these dressed central charges collectively by $\Psi
^{I}\left( \phi \right) $ and express the attractor eqs both in terms of (%
\ref{la}) and (\ref{al}).

\emph{Using eq(\ref{la}) }\newline
By using the matrix fields $L$ and $K$ as well as symmetry under $SO\left(
3\right) \times SO\left( 2\right) $, the extremization of the potential is
given by,%
\begin{equation}
\begin{tabular}{ll}
$\sum \left( \frac{\partial \mathcal{V}_{eff}}{\partial L_{ab}}\right) \frac{%
\partial L_{ab}}{\partial \varphi ^{A}}d\varphi ^{A}+\sum \left( \frac{%
\partial \mathcal{V}_{eff}}{\partial K_{ij}}\right) \frac{\partial K_{ij}}{%
\partial \xi ^{\alpha }}d\xi ^{\alpha }=0$ & .%
\end{tabular}%
\end{equation}%
So the attractor equations read, up to $SO\left( 3\right) \times SO\left(
2\right) $ transformations, as follows:%
\begin{equation}
\begin{tabular}{llll}
$d\varphi ^{A}Tr\left[ L\mathcal{T}_{A}\left( \frac{\partial \mathcal{V}%
_{eff}}{\partial L}\right) \right] =0$ & , & $A=1,\ldots ,8$ &  \\ 
$d\xi ^{\alpha }Tr\left[ K\mathcal{\tau }_{\alpha }\left( \frac{\partial 
\mathcal{V}_{eff}}{\partial K}\right) \right] =0$ & , & $\alpha =0,1,2$ & 
\end{tabular}%
\end{equation}%
in agreement with the $\frac{SL\left( 3,R\right) }{SO\left( 3\right) }\times 
\frac{SL\left( 2,R\right) }{SO\left( 2\right) }$ factorization of the moduli
space.

\emph{Using eq(\ref{al})}\newline
In this case the extremization condition $\delta \mathcal{V}_{eff}=0$ may be
also written as,%
\begin{equation}
\begin{tabular}{l}
$\sum \left( \frac{\partial \mathcal{V}_{eff}}{\partial \Psi ^{I}}\right) 
\frac{\partial \Psi ^{I}}{\partial L_{ab}}\delta L_{ab}+\sum \left( \frac{%
\partial \mathcal{V}_{eff}}{\partial \Psi ^{I}}\right) \frac{\partial \Psi
^{I}}{\partial K_{ab}}\delta K_{ab}=0$%
\end{tabular}
\label{ef}
\end{equation}%
Using eqs(\ref{pot}-\ref{po}), we can bring this constraint relation into
the form,%
\begin{equation}
\begin{tabular}{ll}
$\delta \mathcal{V}_{eff}=$ & $+$ $\sum \delta _{ab}\delta _{ij}\left(
X^{bj}\delta X^{ai}\right) +\sum \delta ^{ab}\delta ^{ij}\left( \tilde{X}%
_{bj}\delta \tilde{X}_{ai}\right) $ \\ 
& $+$ $\sum \delta _{ab}Y^{b}\delta Y^{a}+\sum \delta ^{ab}\tilde{Y}%
_{b}\delta \tilde{Y}_{a}+$ $\sum \delta _{ij}Z^{j}\delta Z^{i}$%
\end{tabular}
\label{xx}
\end{equation}%
with 
\begin{equation}
\begin{tabular}{llllll}
$\delta X^{ai}$ & $=+\sum \left( \Omega _{b}^{a}\delta _{j}^{i}+\omega
_{j}^{i}\delta _{b}^{a}\right) X^{bj}$ & , & $\delta Y^{a}$ & $=+\sum \Omega
_{b}^{a}Y^{b}$ &  \\ 
$\delta \tilde{X}_{ai}$ & $=-\sum \left( \Omega _{a}^{b}\delta
_{i}^{j}+\omega _{i}^{j}\delta _{a}^{b}\right) \tilde{X}_{bj}$ & , & $\delta 
\tilde{Y}_{a}$ & $=-\sum \Omega _{a}^{b}\tilde{Y}_{b}$ &  \\ 
$\delta Z^{i}$ & $=+\sum \omega _{j}^{i}Z^{j}$ & , &  &  & 
\end{tabular}%
\end{equation}%
and where the 1- forms $\Omega =\left( \delta L\right) L^{-1}$ and $\omega
=\left( \delta K\right) K^{-1}$ are respectively the Cartan Maurer forms of
the $SL\left( 3,R\right) $ and $SL\left( 2,R\right) $ that we have studied
previously. Putting these relations back into (\ref{xx}), we can read the
attractor equations from the following relation,%
\begin{equation}
\begin{tabular}{ll}
$\delta \mathcal{V}_{eff}$ & $=Tr\left( X\Omega X-\tilde{X}\Omega \tilde{X}%
\right) +Tr\left( Y\Omega Y-\tilde{Y}\Omega \tilde{Y}\right) $ \\ 
& $+$ $Tr\left( X\omega X-\tilde{X}\omega \tilde{X}\right) +Tr\left( Z\omega
Z\right) =0$%
\end{tabular}%
\end{equation}%
Rewriting this constraint equation as a linear combination of the Maurer
Cartan 1-forms on the $SL\left( 3\right) \otimes SL\left( 2\right) $ group
manifold like 
\begin{equation}
\begin{tabular}{l}
$\sum\limits_{a,b}\Upsilon ^{ba}\Omega _{ab}+\sum\limits_{i,j}\digamma
^{ji}\omega _{ij}=0$%
\end{tabular}
\label{att}
\end{equation}%
with $\Omega _{ab}=\sum_{A}d\varphi ^{A}\left( \Omega _{A}\right) _{ab}$, $%
\omega _{ij}=\sum_{\alpha }d\xi ^{\alpha }\left( \omega _{\alpha }\right)
_{ij}$ and,%
\begin{equation}
\begin{tabular}{ll}
$\Upsilon ^{ab}=\sum \left( \delta _{ij}X^{ai}X^{bj}-\delta ^{ij}\tilde{X}%
_{ai}\tilde{X}_{bj}\right) +\left( Y^{a}Y^{b}-\tilde{Y}_{a}\tilde{Y}%
_{b}\right) $ & , \\ 
$\digamma ^{ij}=\sum \left( \delta _{il}\delta _{ab}X^{al}X^{bj}-\sum \delta
^{ab}\delta ^{ik}\delta ^{jl}\tilde{X}_{ak}\tilde{X}_{bl}\right) +Z^{i}Z^{j}$
& ,%
\end{tabular}
\label{at}
\end{equation}%
the attractor equations can be expressed as follows:%
\begin{equation}
\begin{tabular}{lll}
$\Upsilon ^{ab}$ $\Omega _{\left[ ab\right] }^{{\scriptsize SO}_{%
{\scriptsize 3}}}+\Upsilon ^{ab}$ $\Omega _{\left( ab\right) }^{{\scriptsize %
SL}_{{\scriptsize 3}}{\scriptsize /SO}_{{\scriptsize 3}}}$ & $=0$ & , \\ 
$\digamma ^{ij}$ $\omega _{\left[ ij\right] }^{{\scriptsize SO}_{%
{\scriptsize 2}}}+\digamma ^{ij}$ $\omega _{\left( ij\right) }^{{\scriptsize %
SL}_{{\scriptsize 2}}{\scriptsize /SO}_{{\scriptsize 2}}}$ & $=0$ & ,%
\end{tabular}
\label{a}
\end{equation}%
where the Maurer Cartan 1-forms are as before. Notice that from eqs(\ref{at}%
), the tensors $\Upsilon ^{ab}$ and $\digamma ^{ij}$ are symmetric ($%
\Upsilon ^{ab}=\Upsilon ^{\left( ab\right) }$, $\digamma ^{ij}=\digamma
^{\left( ij\right) }$) and so the first terms of above relations namely $%
\Upsilon ^{\left( ab\right) }$ $\Omega _{\left[ ab\right] }^{{\scriptsize SO}%
_{{\scriptsize 3}}}$ and $\digamma ^{\left( ij\right) }$ $\omega _{\left[ ij%
\right] }^{{\scriptsize SO}_{{\scriptsize 2}}}$ vanish identically. This
property reflets just invariance under the $SO\left( 3\right) \times
SO\left( 2\right) $ isometry of the moduli space. So, the attractor
equations of the black p-branes of maximal supergravity in \emph{8D }read as
follows, 
\begin{equation}
\begin{tabular}{lll}
$\Upsilon ^{\left( ab\right) }\Omega _{\left( ab\right) }^{{\scriptsize SL}_{%
{\scriptsize 3}}{\scriptsize /SO}_{{\scriptsize 3}}}=0$ & , & $\digamma
^{\left( ij\right) }$ $\omega _{\left( ij\right) }^{{\scriptsize SL}_{%
{\scriptsize 2}}{\scriptsize /SO}_{{\scriptsize 2}}}=0$%
\end{tabular}%
\end{equation}%
where now $\Upsilon ^{\left( ab\right) }$ and $\digamma ^{\left( ij\right) }$
are precisely as in eqs(\ref{at}).

\section{Solving the attractor eqs}

We first study the solutions of the attractor eqs for the dyonic black
membrane with the near horizon geometry $\emph{AdS}_{4}\times \emph{S}^{4}$;
actually this completes the partial results given in \textrm{\cite{8A}}.
Then, we examine the other black attractor solutions corresponding to the p-
branes with near horizon geometries $\emph{AdS}_{p+2}\times \emph{S}^{6-p}$.

\subsection{Dyonic membrane}

The dyonic black membrane of $8D$ maximal supergravity has a near horizon
geometry $AdS_{4}\times S^{4}$ in which the electric $e$ and magnetic $g$
charges of the 4-form $\mathcal{F}_{4}$ are switched on%
\begin{equation}
\begin{tabular}{llll}
$\mathcal{F}_{4}=e\alpha _{4}+g\beta _{4}$ & , & $\mathcal{\tilde{F}}%
_{4}\sim \mathcal{F}_{4}$ & .%
\end{tabular}
\label{cl}
\end{equation}%
The real 4-forms $\alpha _{4}=\alpha _{AdS_{4}}$ and $\beta _{4}=\beta
_{S^{4}}$ are respectively the volume forms on the non compact $AdS_{4}$ and
the compact n-sphere $S^{4}$. We have%
\begin{equation}
\begin{tabular}{lll}
$\frac{1}{\mathcal{V}_{AdS_{4}}}\int_{AdS_{4}}^{\Lambda }\alpha _{4}=1$ & ,
& $\frac{1}{\mathcal{V}_{S^{{\scriptsize 4}}}}\int_{S^{4}}\beta _{4}=1$%
\end{tabular}%
\end{equation}%
with $\mathcal{V}_{AdS_{4}}$ describing a regularized volume with $\Lambda $
some UV\ regularization parameter. We also have%
\begin{equation}
\begin{tabular}{ll}
$\int_{{\scriptsize AdS}_{4}{\scriptsize \times S}^{4}}\mathcal{F}_{4}\wedge 
\mathcal{\tilde{F}}_{4}\sim eg\mathcal{V}_{{\scriptsize tot}}$ & ,%
\end{tabular}%
\end{equation}%
where $\mathcal{V}_{{\scriptsize tot}}=\mathcal{V}_{{\scriptsize AdS}%
_{4}}\times \mathcal{V}_{{\scriptsize S}^{{\scriptsize 4}}}$ and where the
electric and magnetic charges of the membrane are related by the Dirac
quantization relation $eg\sim n$. Notice that p-forms with $p\neq 4$, which
are associated with the other p-black branes of the \emph{8D} supergravity,
are not supported by the $AdS_{4}\times S^{4}$ geometry since there is no
p-cycles allowing relations type (\ref{cl}). As such the charges $P^{ai},$ $%
Q_{ia},$ $p^{a},$ $q_{a}$ are switched off; so the attractor eqs reduce to
the following conditions on the dressed central charges, 
\begin{equation}
\begin{tabular}{ll}
$Z^{i}Z^{j}\omega _{ij}=0$ & .%
\end{tabular}
\label{zw}
\end{equation}%
A standard way to deal with relation is to suppose $\omega _{ij}\neq 0$ $%
\forall i,j$; and end with the constraint eqs $Z^{i}Z^{j}=0$ leading to the
trivial solution $Z^{i}=0$; ie $Z_{elc}=Z_{mag}=0$. This solution, which
requires the vanishing of the bare electric and magnetic charges; 
\begin{equation}
g=0\text{ \ \ },\text{ \ \ }e=0,  \label{o}
\end{equation}%
corresponds then to a trivial configuration with no black membrane charges.
To get more insight into eq(\ref{zw}), let us work out explicitly the
minimum of the effective potential $\mathcal{V}_{eff}=\mathcal{V}%
_{eff}\left( \sigma ,\vartheta \right) $ of the black membrane whose
explicit expression is as in eq(\ref{pt}). The corresponding attractor
equations, 
\begin{equation}
\begin{tabular}{llll}
$\frac{\partial \mathcal{V}_{eff}}{\partial \sigma }=0$ & , & $\frac{%
\partial \mathcal{V}_{eff}}{\partial \vartheta }=0$ & ,%
\end{tabular}
\label{c1}
\end{equation}
lead to,%
\begin{equation}
\begin{array}{ccc}
2\mathcal{Z}^{2}\left( Z_{1}Z_{1}-Z_{2}Z_{2}\right) & =0 & , \\ 
2\mathcal{Z}^{2}\left( Z_{1}Z_{2}+Z_{2}Z_{1}\right) & =0 & ,%
\end{array}%
\end{equation}%
where we have set $\mathcal{Z}^{2}=\sum_{i,j}\left( Z^{i}\delta
_{ij}Z^{j}\right) $. Moreover, the Hessian matrix is given by,%
\begin{equation}
\begin{tabular}{lll}
$\frac{\partial ^{2}\mathcal{V}_{eff}}{\partial \sigma ^{2}}$ & $=4\mathcal{Z%
}^{2}\left( Z_{1}Z_{1}-Z_{2}Z_{2}\right) ^{2}$ & , \\ 
$\frac{\partial ^{2}\mathcal{V}_{eff}}{\partial \vartheta ^{2}}$ & $=4%
\mathcal{Z}^{2}\left( Z_{1}Z_{2}+Z_{2}Z_{1}\right) ^{2}$ & , \\ 
$\frac{\partial ^{2}\mathcal{V}_{eff}}{\partial \vartheta \partial \sigma }$
& $=4\mathcal{Z}^{2}\left( Z_{1}Z_{2}+Z_{2}Z_{1}\right) \left(
Z_{1}Z_{1}-Z_{2}Z_{2}\right) $ & .%
\end{tabular}%
\end{equation}%
From these relations, we see that the solution of the attractor eqs (\ref{c1}%
) is given by the trivial values $Z_{1}=0,$ $Z_{2}=0$ requiring $g=0,$ $e=0$
in agreement with eqs(\ref{zw},\ref{o}).

\subsection{Black pairs in $AdS_{2+p}\times S^{6-p}$ geometries}

Here we study the general solutions of the attractor equations concerning
the system made of black strings and their dual magnetic 3-branes in the $%
AdS_{3}\times S^{5}$ geometry. Then, we consider explicit solutions for
black strings recovered from the pair strings/3-branes by switching off the
electric charges $q_{a}$. Similar analysis may be done for the pair black
holes/4-branes; it is omitted.

\subsubsection{black pair: strings/3-branes}

We begin by recalling that the near horizon geometries of the black strings
and the black 3-branes in \emph{8D} maximal supergravity are respectively
given by $AdS_{3}\times S^{5}$ and $AdS_{5}\times S^{3}$. But here we will
mainly focus on the $AdS_{3}\times S^{5}$ geometry; a similar analysis is
also valid for $AdS_{5}\times S^{3}$. Recall that, generally speaking, the
metric of $AdS_{p+2}\times S^{6-p}$ geometry of black p-branes reads as
follows,%
\begin{equation}
ds_{8}^{2}=R_{AdS_{p}}^{2}ds_{AdS_{n}}^{2}+R_{S^{8-n}}^{2}ds_{S^{8-n}}^{2}
\end{equation}%
with%
\begin{equation}
ds_{AdS_{n}}^{2}=d\rho ^{2}-\sinh ^{2}\rho \text{ }d\tau +\cosh \rho \text{ }%
d\varpi _{n-2}^{2}
\end{equation}%
and $\rho \geq 0,$ $\tau \in \left[ 0,2\pi \right] $ and $d\varpi _{n}^{2}$
the length element on the unit n-sphere $S^{n}$ inside the non compact AdS
space. \newline
In the case of the $AdS_{3}\times S^{5}$ near horizon geometry, the black
strings are located in the $AdS_{3}$\ part of the $AdS_{3}\times S^{5}$\
space and their dual magnetic 3-branes wrap the $S^{5}/S^{2}$\ as
illustrated below,%
\begin{eqnarray}
&&%
\begin{tabular}{l|l|l|l|l|l|l|l|l|lll|}
\cline{2-12}
& \multicolumn{3}{|l|}{$AdS_{3}$} & \multicolumn{5}{|l|}{$\ \ \ \ \ \ \ \
S^{5}$} & \multicolumn{3}{|l|}{\ \ \ $T^{3}$} \\ \cline{2-12}
& 0 & 1 & 2 & 3 & 4 & 5 & 6 & 7 & 8 & 9 & 10 \\ \hline
\multicolumn{1}{|l|}{M2} & $\mathbf{\times }$ & $\mathbf{\times }$ & $%
\mathbf{\circ }$ & $\mathbf{\circ }$ & $\mathbf{\circ }$ & $\mathbf{\circ }$
& $\mathbf{\circ }$ & $\mathbf{\circ }$ & $\mathbf{\times }$ & $\mathbf{%
\circ }$ & $\mathbf{\circ }$ \\ \hline
\multicolumn{1}{|l|}{M5} & $\mathbf{\times }$ & $\mathbf{\circ }$ & $\mathbf{%
\circ }$ & $\mathbf{\times }$ & $\mathbf{\times }$ & $\mathbf{\times }$ & $%
\mathbf{\circ }$ & $\mathbf{\circ }$ & $\mathbf{\circ }$ & $\mathbf{\times }$
& $\mathbf{\times }$ \\ \hline
\end{tabular}
\\
&&  \notag
\end{eqnarray}%
together with the two other permutations in the 3-torus directions. In this
view, the field strengths $\mathcal{F}_{3}^{a}$ that couple the strings and
their magnetic duals $\mathcal{\tilde{F}}_{5a}$ that couple the 3-branes are
respectively given by $\mathcal{F}_{3}^{a}=p^{a}\alpha _{3}$ and $\mathcal{%
\tilde{F}}_{5a}=q_{a}\beta _{5}$. The magnetic charges $p^{a}$ of the
strings and the electric $q_{a}$ of the 3- branes are given by the fluxes, 
\begin{equation}
\begin{tabular}{llll}
$p^{a}\sim $ & $\int_{AdS_{3}}\ \mathcal{F}_{3}^{a}$ & $=$ & $p^{a}$ $%
\int_{AdS_{3}}\ \alpha _{3}$, \\ 
$q_{a}\sim $ & $\int_{S^{5}}$ $\ \mathcal{\tilde{F}}_{5a}$ & $=$ & $q_{a}$ $%
\int_{S^{5}}\ \beta _{5}$,%
\end{tabular}%
\end{equation}%
where $\alpha _{3}$ and $\beta _{5}$ stand for the volume forms $\alpha
_{3}^{{\scriptsize AdS}_{3}}$ and $\beta _{5}^{{\scriptsize S}^{{\scriptsize %
5}}}$. We also have,%
\begin{equation}
\begin{tabular}{ll}
$\int_{AdS_{3}\times S^{5}}\mathcal{F}_{3}^{a}\wedge \mathcal{\tilde{F}}%
_{5b}\sim p^{a}q_{b}\mathcal{V}_{AdS_{3}\times S^{5}}$ & ,%
\end{tabular}%
\end{equation}%
where $\mathcal{V}_{AdS_{3}\times S^{5}}=\mathcal{V}_{AdS_{3}}\times 
\mathcal{V}_{S^{5}}$.\newline
The attractor equations describing the system of black strings/3-branes in
the $AdS_{3}\times S^{5}$ geometry follow from the extremization of their
effective potential $\mathcal{V}_{1}+\mathcal{V}_{3}$ $=$ $\mathcal{V}_{1}+%
\mathcal{\tilde{V}}_{1}$ which reads in terms of the dressed charges $Y^{a}$%
, $\tilde{Y}_{a}$ as follows 
\begin{equation}
\begin{tabular}{ll}
$\mathcal{V}_{1}+\mathcal{\tilde{V}}_{1}=\frac{1}{2}\sum Y^{a}\delta
_{ab}Y^{b}+\frac{1}{2}\sum \tilde{Y}_{a}\delta ^{ab}\tilde{Y}_{b}$ & $.$%
\end{tabular}%
\end{equation}%
The computation of $\delta \left( \mathcal{V}_{1}+\mathcal{\tilde{V}}%
_{1}\right) =0$ leads to the condition $\left( Y^{a}Y^{b}-\tilde{Y}^{a}%
\tilde{Y}^{b}\right) \Omega _{\left( ab\right) }=0$ where the $3\times 3$
matrix field $\Omega $ is the Maurer Cartan 1-form on $SL\left( 3,R\right) $%
. Now, seen that $\left( Y^{a}Y^{b}-\tilde{Y}^{a}\tilde{Y}^{b}\right) $ is a
symmetric matrix, the above condition reduces to, 
\begin{equation}
\begin{tabular}{ll}
$\left( Y^{a}Y^{b}-\tilde{Y}^{a}\tilde{Y}^{b}\right) \Omega _{ab}^{SL_{%
{\scriptsize 3}}/SO_{{\scriptsize 3}}}=0$ & ,%
\end{tabular}
\label{yy}
\end{equation}%
where now $\Omega _{ab}^{SL_{{\scriptsize 3}}/SO_{{\scriptsize 3}}}$ is the
Maurer Cartan 1-forms on the $SL\left( 3,R\right) /SO\left( 3\right) $ coset
manifold, in agreement with $SO\left( 3\right) \times SO\left( 2\right) $
isometry of the moduli space. The attractor equations reads then as follows:%
\begin{equation}
\begin{tabular}{llll}
$\Omega _{ab}^{SL_{{\scriptsize 3}}/SO_{{\scriptsize 3}}}\neq 0$ & , & $%
\left( Y^{a}Y^{b}-\tilde{Y}^{a}\tilde{Y}^{b}\right) =0$ & $.$%
\end{tabular}
\label{y}
\end{equation}%
Since the dressed charges $Y^{a}$, $\tilde{Y}^{a}$ depend on the matrix
field $L$ and the charges $p^{a}$, $q_{a}$, the solving of the above
equations turns to fixing the fields $L_{ab}$ in terms of the electric and
magnetic charges of the black objects. A particular solution with $p\neq 0$, 
$q\neq 0$ is given by 
\begin{equation}
\begin{tabular}{llll}
$\left( L^{2}\right) _{ba}\sim \frac{p_{b}q_{a}}{p^{2}}$ & , & $\left(
L^{-2}\right) ^{ab}\sim \frac{q^{a}p^{b}}{q^{2}}$ & ,%
\end{tabular}%
\end{equation}%
with $p^{2}=\sum p_{c}p^{c}$ and $q^{2}=\sum q_{c}q^{c}$. \newline
Notice that\textrm{\ expanding the real 1-form matrix field }$\Omega $%
\textrm{\ along the basis of }$SL\left( 3,R\right) $\textrm{\ as follows, }%
\begin{equation}
\begin{tabular}{ll}
$\Omega =\sum\limits_{A=1}^{8}\mathcal{T}^{A}\text{ }\Delta _{A}$ & ,%
\end{tabular}
\label{dc}
\end{equation}%
we can also put (\ref{yy}) in the equivalent form%
\begin{equation}
\begin{tabular}{ll}
$\sum\limits_{A=1}^{8}\text{ }\Delta _{A}\sum \left( Y^{a}\mathcal{T}%
_{ab}^{A}Y^{b}-\tilde{Y}^{a}\mathcal{T}_{ab}^{A}\tilde{Y}^{b}\right) =0$ & ,%
\end{tabular}%
\end{equation}%
In terms of these matrices, the attractor eqs read as follows, 
\begin{equation}
\begin{tabular}{llll}
$\left( Y\mathcal{T}^{A}Y\right) -\left( \tilde{Y}\mathcal{T}^{A}\tilde{Y}%
\right) =0$ & , & $\Delta _{A}\neq 0$ & .%
\end{tabular}%
\end{equation}%
Notice the three following:\newline
\textbf{(i)} the attractor eqs associted with the SO$\left( 3\right) $
geberators given by values $A=1,2,3$ vanish identically since for these
values we have $\mathcal{T}_{ab}^{A}=-\mathcal{T}_{ba}^{A}$.\ \newline
\textbf{(ii)} the non zero contributions comes from the remaining five
generators $\mathcal{T}_{4},$..., $\mathcal{T}_{8}$. These $3\times 3$
generators are real, symmetric and traceless matrices.\newline
\textbf{(iii)} Altogether with the $SO\left( 3\right) $ generators $\mathcal{%
T}_{1}$, $\mathcal{T}_{2}$, $\mathcal{T}_{3}$ given by (\ref{3g}), the real
matrices $\mathcal{T}_{4},$..., $\mathcal{T}_{8}$ generate the eight
dimensional $SL\left( 3,R\right) $ symmetry. These $\mathcal{T}_{4},$..., $%
\mathcal{T}_{8}$ matrices can be explicitly read by help of eq(\ref{f});
from which we learn that the two diagonal generators $T_{4}$ and $T_{5}$ are
given by 
\begin{equation}
\begin{tabular}{ll}
$\mathcal{T}_{4}=\left( 
\begin{array}{ccc}
1 & 0 & 0 \\ 
0 & 0 & 0 \\ 
0 & 0 & -1%
\end{array}%
\right) ,$ & $\mathcal{T}_{5}=\left( 
\begin{array}{ccc}
0 & 0 & 0 \\ 
0 & 1 & 0 \\ 
0 & 0 & -1%
\end{array}%
\right) $%
\end{tabular}
\label{ta}
\end{equation}%
and the three non diagonal ones are as follows%
\begin{equation}
\begin{tabular}{lll}
$\mathcal{T}_{6}=\left( 
\begin{array}{ccc}
0 & 1 & 0 \\ 
1 & 0 & 0 \\ 
0 & 0 & 0%
\end{array}%
\right) ,$ & $\mathcal{T}_{7}=\left( 
\begin{array}{ccc}
0 & 0 & 1 \\ 
0 & 0 & 0 \\ 
1 & 0 & 0%
\end{array}%
\right) ,$ & $\mathcal{T}_{8}=\left( 
\begin{array}{ccc}
0 & 0 & 0 \\ 
0 & 0 & 1 \\ 
0 & 1 & 0%
\end{array}%
\right) $%
\end{tabular}
\label{tb}
\end{equation}%
Putting the decomposition (\ref{dc}) back into eq(\ref{yy}), we can bring it
to the form $\sum_{A=4}^{8}\Upsilon ^{A-3}$ $\Delta _{A}=0$, with $\Upsilon
^{A}=Y\mathcal{T}^{A}Y-\tilde{Y}\mathcal{T}^{A}\tilde{Y}$ or more explicitly 
\begin{equation}
\Upsilon ^{A}=\sum_{a,b=1}^{3}\left( Y^{a}Y^{b}-\tilde{Y}^{a}\tilde{Y}%
^{b}\right) \mathcal{T}_{ab}^{A}\text{ }.  \label{aw}
\end{equation}%
By substituting $\mathcal{T}_{ab}^{A}$ by their values (\ref{ta}), we can
also put the components $\Upsilon ^{A}$ like,%
\begin{equation}
\begin{tabular}{ll}
$\Upsilon ^{1}$ & $=\left( Y^{1}Y^{1}-Y^{3}Y^{3}\right) -\left( \tilde{Y}^{1}%
\tilde{Y}^{1}-\tilde{Y}^{3}\tilde{Y}^{3}\right) $ \\ 
$\Upsilon ^{2}$ & $=\left( Y^{2}Y^{2}-Y^{3}Y^{3}\right) -\left( \tilde{Y}^{2}%
\tilde{Y}^{2}-\tilde{Y}^{3}\tilde{Y}^{3}\right) $ \\ 
$\Upsilon ^{3}$ & $=2\left( Y^{1}Y^{2}-\tilde{Y}^{1}\tilde{Y}^{2}\right) $
\\ 
$\Upsilon ^{4}$ & $=2\left( Y^{1}Y^{3}-\tilde{Y}^{1}\tilde{Y}^{3}\right) $
\\ 
$\Upsilon ^{5}$ & $=2\left( Y^{2}Y^{3}-\tilde{Y}^{2}\tilde{Y}^{3}\right) $%
\end{tabular}
\label{tt}
\end{equation}%
where $Y^{a}$ \ and $\tilde{Y}^{a}$\ are as in eqs(\ref{ya}). Combining eqs(%
\ref{w}-\ref{aw}), we learn that, depending on the values of dressed charges
Y and \~{Y}, there are several kinds of solutions for the attractor equations%
\begin{equation}
\Upsilon ^{1}=\Upsilon ^{2}=\Upsilon ^{3}=\Upsilon ^{4}=\Upsilon ^{5}=0\text{%
,}
\end{equation}%
whose solution reads, up to $SO\left( 3\right) $ transformation, as follows, 
\begin{equation}
\begin{tabular}{lll}
$Y^{a}=\left( 
\begin{array}{c}
y \\ 
0 \\ 
0%
\end{array}%
\right) $ & , & $\tilde{Y}^{a}=\pm \left( 
\begin{array}{c}
y \\ 
0 \\ 
0%
\end{array}%
\right) $%
\end{tabular}
\label{2s}
\end{equation}%
with $y$ is a non zero real number. These relations describe two solutions;
one with a sign $\left( +\right) $ corresponding to a black string/3-brane
pair and the second with sign $\left( -\right) $ associated with a black
string/anti 3- brane pair. These solutions fix three real scalars amongst
the seven ones parameterizing $\frac{SL\left( 3,R\right) \times SL\left(
2,R\right) }{SO\left( 3\right) \times SO\left( 2\right) }$ reducing thus the
moduli space down to 
\begin{equation*}
\frac{SL\left( 2,R\right) \times SL\left( 2,R\right) }{SO\left( 2\right)
\times SO\left( 2\right) }.
\end{equation*}

\subsubsection{black strings}

The attractor equations for the black strings may be obtained by starting
from eqs(\ref{tt}) describing the attractor eqs of the strings/ (anti)
3-brane pairs; then set to zero the charges of the (anti) 3-branes 
\begin{equation}
q_{a}=\tilde{Y}_{a}=0.
\end{equation}%
This leads to the relations%
\begin{equation}
\begin{tabular}{ll}
$\theta _{1}\left( Y^{1}Y^{1}-Y^{3}Y^{3}\right) $ & $=0$ \\ 
$\theta _{2}\left( Y^{2}Y^{2}-Y^{3}Y^{3}\right) $ & $=0$ \\ 
$\theta _{3}Y^{1}Y^{2}$ & $=0$ \\ 
$\theta _{4}Y^{1}Y^{3}$ & $=0$ \\ 
$\theta _{5}Y^{2}Y^{3}$ & $=0$%
\end{tabular}
\label{ae}
\end{equation}%
where $Y^{a}$ are the dressed charges associated with the black strings and
where we have set $\theta _{A}=\Delta _{A-4}$. In the case where all the $%
\Delta _{A}$'s are non zero, it is clear that all the dressed charges should
vanish 
\begin{equation}
\begin{tabular}{llll}
$Y^{a}=0$ & , & $\theta _{A}\neq 0$ & $.$%
\end{tabular}%
\end{equation}

\section{Intersecting attractors}

From the results of \textrm{\cite{8A}, }we learn that one should distinguish
two main classes of black p-brane solutions in higher dimensional
supergravity. In the \emph{8D} case we are interested in here, these are: 
\newline
(\textbf{1}) the standard black p- brane solutions based on $AdS_{2+p}\times
S^{6-p}$, $p=0,1,2,3,4$, whose features have been explicitly analyzed in
previous sections.\newline
(\textbf{2}) the so called intersecting attractors with the typical near
horizon geometries 
\begin{equation}
\begin{tabular}{lllll}
$AdS_{2+p}\times S^{m}\times T^{6-p-m}$ & , & $p=0,1,2,3,4$ & , & $p+m\leq 5$%
.%
\end{tabular}
\label{ad}
\end{equation}%
The novelty with these geometries is that they allow the two following: (%
\textbf{i}) a variety of irreducible sub-manifolds that support various
kinds of branes and so a rich spectrum of electric and magnetic charges; (%
\textbf{ii}) non trivial intersections between $p_{i}$-/$p_{j}$- cycles of (%
\ref{ad}) leading to intersecting (BPS and non BPS) attractors. To
illustrate the first point, we consider the example of the two compact
manifolds $S^{m+n}$ and $\mathcal{M}^{m+n}=S^{m}\times T^{n}$ with same
dimension. While the sphere $S^{m+n}$ supports only charges of $\left(
m+n-2\right) $-brane charges%
\begin{equation}
\begin{tabular}{llll}
$\mathcal{F}_{n+m}=g$ $\varpi _{n+m}$ & , & $g=\int_{S^{m+n}}\mathcal{F}%
_{n+m}$ & ,%
\end{tabular}%
\end{equation}%
and no $\left( m-1\right) $- brane nor others, the manifold $S^{m}\times
T^{n}$ allows however many possibilities. It has several irreducible $p_{i}$%
- cycles that support, in addition to $\left( m+n-2\right) $-branes, other
kinds; in particular $n$ types of $\left( m-1\right) $-branes with charges
given by,%
\begin{equation}
\begin{tabular}{llll}
$g^{a}=\int_{\mathcal{C}_{m+1}^{(a)}}\mathcal{F}_{m+1}$ & $,$ & $\mathcal{F}%
_{m+1}=\sum\limits_{a}g^{a}\varpi _{_{m+1|a}}$ & , \\ 
$\int_{\mathcal{C}_{m+1}^{(a)}}\varpi _{_{m+1|b}}=\delta _{b}^{a}$ & , & $%
a=1,...,n$ & ,%
\end{tabular}%
\end{equation}%
where the $\mathcal{C}_{m+1}^{(a)}$ cycles stand for $\bigcup_{a=1}^{n}%
\left( S_{{\scriptsize a}}^{1}\times S^{m}\right) $ with $\prod_{a=1}^{n}S_{%
{\scriptsize a}}^{1}=T^{n}$. The branes may be imagined as filling the fiber 
$\mathsf{F}_{m-1}^{(a)}$ of these cycles $\mathcal{C}_{m+1}^{(a)}$ thought
of in terms of the fibration $\mathcal{C}_{m+1}^{(a)}\sim \mathsf{F}%
_{m-1}^{(a)}\times S^{2}$ with field strength $\mathcal{F}_{m+1}=\beta
_{S^{2}}\wedge \left( \sum\limits_{a}g^{a}\beta _{\mathsf{F}%
_{m-1}^{(a)}}\right) $.\newline
Using the anzats of \textrm{\cite{8A}}, we focus below on the study of
various examples of these typical horizon geometries and work out new and
explicit solutions regarding intersecting attractors in the case of \emph{8D}
maximal supergravity. As the solutions are very technical, we will
concentrate on drawing the crucial lines and give the results.

\subsection{Geometries with AdS$_{4}$ factor}

We consider two examples: (\textbf{a}) \emph{AdS}$_{4}\times $\emph{S}$%
^{3}\times $\emph{S}$^{1}$ and (\textbf{b}) \emph{AdS}$_{4}\times $\emph{S}$%
^{2}\times $\emph{T}$^{2}$; the other possibility namely \emph{AdS}$%
_{4}\times $\emph{S}$^{4}$ has been considered in subsection 4.1. On the 
\emph{AdS}$_{4}\times $\emph{S}$^{3}\times $\emph{S}$^{1}$ near horizon
geometry, the non vanishing field strength charges of the \emph{8D} maximal
supergravity are: (\textbf{i}) the magnetic $p^{a}$ of the strings, (\textbf{%
ii) }the $q_{a}$ of the 3-branes and (\textbf{iii) }the $\left( e,g\right) $
charge of the dyonic membrane. In the case of \emph{AdS}$_{4}\times $\emph{S}%
$^{2}\times $\emph{T}$^{2}$, there are moreover non trivial black hole
charges and 4-brane ones; more precisely:

\emph{Field strengths on} \emph{AdS}$_{4}\times $\emph{S}$^{3}\times $\emph{S%
}$^{1}$\newline
Using the volume forms $\alpha _{AdS_{4}}$ and $\beta _{S^{n}}$ with $n=1,3$%
, we have the following field strengths relations:

\begin{equation}
\begin{tabular}{l}
$%
\begin{tabular}{l|l|l|}
\cline{2-3}
& p-branes & $\left( 4-p\right) $- branes \\ \hline
\multicolumn{1}{|l|}{$p=0$} & $\mathcal{F}_{2}^{ai}=0$ & $\mathcal{\tilde{F}}%
_{6|ai}=0$ \\ \hline
\multicolumn{1}{|l|}{$p=1$} & $\mathcal{F}_{3}^{a}=p^{a}\beta _{S^{3}}$ & $%
\mathcal{\tilde{F}}_{5|a}=q_{a}\alpha _{AdS_{4}}{\small \wedge }\beta
_{S^{1}}$ \\ \hline
\multicolumn{1}{|l|}{$p=2$} & \multicolumn{2}{|l|}{$\ \ \ \ \mathcal{F}_{4}=e%
\text{ }\alpha _{AdS_{4}}+g\left( \beta _{S^{3}}{\small \wedge }\beta
_{S^{1}}\right) $} \\ \hline
\end{tabular}%
$ \\ 
\end{tabular}%
\end{equation}%
The vanishing of the charges of the fields strengths $\mathcal{F}_{2}^{ai}$
and $\mathcal{\tilde{F}}_{6|ai}$ are due to the fact that there is no
compact 2-cycles nor 6-cycles on the \emph{AdS}$_{4}\times $\emph{S}$%
^{3}\times $\emph{S}$^{1}$ geometry that support black holes and black
4-brane charges.

\emph{Field strengths on AdS}$_{4}\times $\emph{S}$^{2}\times $\emph{T}$^{2}$%
\newline
This horizon geometry has, in addition to non compact \emph{AdS}$_{4}$
thought of as a regularized 4-cycle $\int_{AdS_{4}}^{\Lambda }\alpha
_{AdS_{4}}=\mathcal{V}_{_{AdS_{4}}}\left( \Lambda \right) $ with
regularization parameter $\Lambda $, compact n-cycles $\mathcal{C}_{n}^{%
{\small (x)}}\subseteq $ \emph{S}$^{2}\times $\emph{T}$^{2}$, with $%
n=1,2,3,4 $ and regularized m-cycles $\mathcal{R}_{m}=$\emph{AdS}$_{4}\times 
\mathcal{C}_{n}^{{\small (x)}}$ with $m=n+4$ supporting brane charges. Using
the anzats of \textrm{\cite{8A}}, we have the corresponding field strengths:%
\begin{equation}
\begin{tabular}{l}
$%
\begin{tabular}{l|l|l|}
\cline{2-3}
& p-branes & $\left( 4-p\right) $- branes \\ \hline
\multicolumn{1}{|l|}{$p=0$} & $\mathcal{F}_{2}^{ai}=P^{ai}\beta _{S^{2}}$ & $%
\mathcal{\tilde{F}}_{6|ai}=Q_{ai}\alpha _{AdS_{4}}{\small \wedge }\beta
_{T^{2}}$ \\ \hline
\multicolumn{1}{|l|}{$p=1$} & $\mathcal{F}_{3}^{a}=\sum\limits_{r=1}^{2}%
\text{ }p_{r}^{a}\beta _{S^{2}}{\small \wedge }\beta _{S_{r}^{1}}$ & $%
\mathcal{\tilde{F}}_{5|a}=\sum\limits_{r}q_{ar}\alpha _{AdS_{4}}{\small %
\wedge }\beta _{\tilde{S}_{r}^{1}}$ \\ \hline
\multicolumn{1}{|l|}{$p=2$} & \multicolumn{2}{|l|}{$\ \ \ \ \mathcal{F}_{4}=e%
\text{ }\alpha _{AdS_{4}}+g\left( \beta _{S^{2}}{\small \wedge }\beta
_{T^{2}}\right) $} \\ \hline
\end{tabular}%
$ \\ 
\end{tabular}
\label{22}
\end{equation}%
where the black string and the black 3-brane fill respectively the $S^{1}$
and $\tilde{S}^{1}$ cycles in the 2-torus $T^{2}=S^{1}\times \tilde{S}^{1}$.
With these electric and magnetic bare charges, we can deduce the dressed
ones and derive the effective potentials associated with these
configurations.

\subsubsection{\emph{AdS}$_{4}\times $\emph{S}$^{3}\times $\emph{S}$^{1}$}

In the case of \emph{AdS}$_{4}\times $\emph{S}$^{3}\times $\emph{S}$^{1}$
geometry, the corresponding effective potential $\mathcal{V}_{eff}$ can be
read from (\ref{pot},\ref{po},\ref{pt}) namely $\frac{1}{2}Y^{a}\delta
_{ab}Y^{b}+\frac{1}{2}\tilde{Y}_{a}\delta ^{ab}\tilde{Y}_{b}+\frac{1}{2}%
Z^{i}\delta _{ij}Z^{j}$; its extremization $\delta \mathcal{V}_{eff}=0$
gives 
\begin{equation}
\begin{tabular}{ll}
$\sum\limits_{A}\Delta _{A}\left( Y\mathcal{T}^{A}Y-\tilde{Y}\mathcal{T}^{A}%
\tilde{Y}\right) \text{ }+\sum\limits_{\alpha }\lambda _{\alpha }$ $\left(
Z\tau ^{\alpha }Z\right) \text{ }=0$ & ,%
\end{tabular}
\label{aa}
\end{equation}%
where $\Delta _{A}$ and $\lambda _{\alpha }$ are 1-forms on the moduli space
as in (\ref{fo}) and where $\mathcal{T}^{A}$, $\tau ^{\alpha }$ are
respectively the generators of SL$\left( 3,R\right) $ and SL$\left(
2,R\right) $. The attractor eqs following from the above extremum namely 
\begin{equation}
\begin{tabular}{lll}
$Y\mathcal{T}^{A}Y-\tilde{Y}\mathcal{T}^{A}\tilde{Y}$ & $=0$ & , \\ 
$Z\tau ^{\alpha }Z$ & $=0$ & ,%
\end{tabular}%
\end{equation}%
are solved as $Y^{a}=+\tilde{Y}^{a}$ (or $Y^{a}=-\tilde{Y}^{a}$ ) and $%
Z_{elec}=Z_{mag}=0$. This configuration describes three dual pairs of
strings/3-branes (or strings/anti- 3-branes) intersecting along the time
direction and has a nice interpretation in M-theory compactified on $T^{3}$.
A typical configuration involving one string and one 3-brane is given by the
following wrapped M2/M5 system%
\begin{eqnarray}
&&%
\begin{tabular}{l|l|l|l|l|l|l|l|l||lll|}
\cline{2-12}
& \multicolumn{4}{|l}{$AdS_{4}$} & \multicolumn{3}{|l}{$\ \ \ \ \ S^{3}$} & $%
S^{1}$ & \multicolumn{3}{||l|}{\ \ \ $T^{3}$} \\ \cline{2-12}
& 0 & 1 & 2 & 3 & 4 & 5 & 6 & 7 & 8 & 9 & 10 \\ \hline
\multicolumn{1}{|l|}{M2} & $\times $ & $\times $ & $\mathbf{\circ }$ & $%
\mathbf{\circ }$ & $\mathbf{\circ }$ & $\mathbf{\circ }$ & $\mathbf{\circ }$
& $\mathbf{\circ }$ & $\mathbf{\times }$ & $\mathbf{\circ }$ & $\mathbf{%
\circ }$ \\ \hline
\multicolumn{1}{|l|}{M5} & $\times $ & $\mathbf{\circ }$ & $\mathbf{\circ }$
& $\mathbf{\circ }$ & $\times $ & $\times $ & $\times $ & $\mathbf{\circ }$
& $\mathbf{\circ }$ & $\mathbf{\times }$ & $\mathbf{\times }$ \\ \hline
\end{tabular}
\\
&&  \notag
\end{eqnarray}%
The two other possible configurations correspond to permuting the role of
the coordinates of the 3-torus.\newline

\subsubsection{\emph{AdS}$_{4}\times $\emph{S}$^{2}\times $\emph{T}$^{2}$
case}

On the geometry \emph{AdS}$_{4}\times $\emph{S}$^{2}\times $\emph{T}$^{2}$
involving the volume forms $\alpha _{AdS_{4}}$, $\beta _{S^{2}}$, $\beta
_{S_{r}^{1}}$, the non vanishing field strength charges are given by eq(\ref%
{22}). Using same approach as before, we can determine the effective
potential $\mathcal{V}_{eff}$ associated with this configuration; its Its
extremization $\delta \mathcal{V}_{eff}=0$ leads to%
\begin{equation}
\begin{tabular}{ll}
$\sum\limits_{A}\Delta _{A}\sum\limits_{i=1}^{2}\left( X^{ai}\mathcal{T}%
_{ab}^{A}X^{bi}-\tilde{X}^{ai}\mathcal{T}_{ab}^{A}\tilde{X}^{bi}+Y^{ai}%
\mathcal{T}_{ab}^{A}Y^{bi}-\tilde{Y}^{ai}\mathcal{T}_{ab}^{A}\tilde{Y}%
^{bi}\right) $ &  \\ 
$+\sum\limits_{\alpha }\lambda _{\alpha }$ $\left( \delta _{ab}\left[
X^{ai}\tau _{ij}^{\alpha }X^{bj}-\tilde{X}^{ai}\tau _{ij}^{\alpha }\tilde{X}%
^{bj}\right] +Z^{i}\tau _{ij}^{\alpha }Z^{i}\right) $ $=0$ & ,%
\end{tabular}%
\end{equation}%
where $\Delta _{A}$, $\lambda _{\alpha }$ $\mathcal{T}^{A}$, $\tau ^{\alpha
} $ are same as before. The attractor eqs,%
\begin{equation}
\begin{tabular}{lll}
$\sum\limits_{i=1}^{2}\left( X^{ai}\mathcal{T}_{ab}^{A}X^{bi}-\tilde{X}^{ai}%
\mathcal{T}_{ab}^{A}\tilde{X}^{bi}+Y^{ai}\mathcal{T}_{ab}^{A}Y^{bi}-\tilde{Y}%
^{ai}\mathcal{T}_{ab}^{A}\tilde{Y}^{bi}\right) $ & $=0$ & , \\ 
$\delta _{ab}\left( X^{ai}\tau _{ij}^{\alpha }X^{bj}-\tilde{X}^{ai}\tau
_{ij}^{\alpha }\tilde{X}^{bj}\right) +Z^{i}\tau _{ij}^{\alpha }Z^{i}$ & $=0$
& ,%
\end{tabular}
\label{fir}
\end{equation}
have two classes of solutions depending on whether $Z^{i}=0$ or $Z^{i}\neq 0$%
. We have:

\emph{Case }$\left( Z_{elec},Z_{mag}\right) =\left( 0,0\right) $\newline
In this case, the corresponding attractor eqs may be solved in various ways;
in particular by compensating the terms of the sum like $X^{ai}=\pm \tilde{X}%
^{ai}$, $Y^{ai}=\pm \tilde{Y}^{ai}$. These configurations describe
intersecting attractors involving black holes, black strings and their duals.

\emph{Case }$\left( Z_{elec},Z_{mag}\right) \neq \left( 0,0\right) $\newline
One of the solutions of the attractor eqs (\ref{fir}) consists to compensate
the terms of the sum as follows: \newline
First solve the first relation of (\ref{fir}) like $X^{ai}=\pm \tilde{Y}%
^{ai} $, $Y^{ai}=\pm \tilde{X}^{ai}$, \newline
Then, solve the second relation by taking $X^{ai}=\pm \nu \tilde{X}^{ai}$
with $\nu $\ some real number; this leads to%
\begin{equation}
\left[ \left( \nu ^{2}-1\right) \tilde{X}^{ai}\tau _{ij}^{\alpha }\tilde{X}%
^{bj}\delta _{ab}+Z^{i}\tau _{ij}^{\alpha }Z^{i}\right] =0
\end{equation}%
from which we learn 
\begin{equation}
Z^{i}=\pm \sqrt{\frac{1-\nu ^{2}}{3}}\sum_{a=1}^{3}\tilde{X}^{ai}  \label{12}
\end{equation}%
where reality property of the central charges $Z^{i}$ imposes to the free
parameter $\nu $ to belong to the set $\left[ 0,1\right] $.

\subsection{Geometries with AdS$_{3}$}

We distinguish two cases: \emph{AdS}$_{3}\times $\emph{S}$^{3}\times $\emph{T%
}$^{2}$ and \emph{AdS}$_{3}\times $\emph{S}$^{2}\times $\emph{T}$^{3}$; here
we focus on the first case as it allows more possibilities. The fluxes
emanating from the black branes associated with \emph{AdS}$_{3}\times $\emph{%
S}$^{3}\times $\emph{T}$^{2}$ are given by,%
\begin{equation}
\begin{tabular}{l|l|l|}
\cline{2-3}
& p-branes & $\left( 4-p\right) $- branes \\ \hline
\multicolumn{1}{|l|}{$p=0$} & $\mathcal{F}_{2}^{ai}=0$ & $\mathcal{\tilde{F}}%
_{6|ai}=0$ \\ \hline
\multicolumn{1}{|l|}{$p=1$} & $\mathcal{F}_{3}^{a}=p^{a}{\Large \beta }%
_{S^{3}}$ & $\mathcal{\tilde{F}}_{5|a}=q^{a}{\Large \alpha }_{AdS_{3}}%
{\small \wedge }$ $\beta _{T^{2}}$ \\ \hline
\multicolumn{1}{|l|}{$p=2$} & \multicolumn{2}{|l|}{$\ \ \ \ \mathcal{F}%
_{4}=\sum\limits_{r=1,2}e_{r}{\Large \alpha }_{AdS_{3}}\wedge {\Large \alpha 
}_{S_{r}^{1}}+\sum\limits_{r=1,2}g_{r}{\Large \beta }_{S_{3}}{\scriptsize %
\wedge }$ ${\Large \beta }_{S_{r}^{1}}$} \\ \hline
\end{tabular}%
\end{equation}%
from which we read the total effective potential,%
\begin{equation}
\mathcal{V}_{eff}=\frac{1}{2}\sum\limits_{r=1}^{2}\left(
Z_{el,r}^{2}+Z_{mag,r}^{2}\right) +\frac{1}{2}\left( Y^{ai}\delta
_{ab}Y^{bi}+\tilde{Y}^{aj}\delta _{ab}\tilde{Y}^{bj}\right)
\end{equation}%
and whose extremization of $\mathcal{V}_{eff}$ gives,%
\begin{equation*}
\sum_{A}\Delta _{A}\left( Y\mathcal{T}^{A}Y-\tilde{Y}\mathcal{T}^{A}\tilde{Y}%
\right) +\sum_{\alpha }\lambda _{\alpha }Z\tau ^{\alpha }Z=0.
\end{equation*}%
The corresponding attractor equations are given by%
\begin{equation}
\begin{tabular}{lllll}
$\sum\limits_{a,b}\left( Y^{a}\mathcal{T}_{ab}^{A}Y^{b}-\tilde{Y}^{a}%
\mathcal{T}_{ab}^{A}\tilde{Y}^{b}\right) $ & $=0$ & , & $A=1,...,8$ &  \\ 
$\sum\limits_{i,j,r}\left( Z_{r}^{i}\tau _{ij}^{\alpha }Z_{r}^{j}\right) $ & 
$=0$ & , & $\alpha =1,2,3$ & 
\end{tabular}%
\end{equation}%
The first relations are solved as usual that is $Y^{ai}=\pm \tilde{Y}^{ai}$
and the second ones like $Z_{r}^{i}=z\delta _{r}^{i}$; thanks to the
identity $z^{2}Tr\left( \tau ^{\alpha }\right) =0$. This configuration
describes an intersecting attractor made of black string/black 3-branes and
black membrane,%
\begin{equation}
\begin{tabular}{l}
$%
\begin{tabular}{l|l|l|l|l|l|l|l|l||lll|}
\cline{2-12}
& \multicolumn{3}{|l}{\ \ \emph{AdS}$_{3}$} & \multicolumn{3}{|l}{\ \ \ 
\emph{S}$^{3}$} & \multicolumn{2}{|l}{\emph{T}$^{2}$} & 
\multicolumn{3}{||l|}{\ \ \ $\ \ \ \ \ \ \ \ T^{3}$} \\ \cline{2-12}
& 0 & 1 & 2 & 3 & 4 & 5 & 6 & 7 & 8 & 9 & 10 \\ \hline
\multicolumn{1}{|l|}{M2} & $\times $ & $\times $ & $\mathbf{\times }$ & $%
\mathbf{\circ }$ & $\mathbf{\circ }$ & $\mathbf{\circ }$ & $\mathbf{\circ }$
& $\mathbf{\circ }$ & $\mathbf{\circ }$ & $\mathbf{\circ }$ & $\mathbf{\circ 
}$ \\ \hline
\multicolumn{1}{|l|}{M2} & $\times $ & $\times $ & $\mathbf{\circ }$ & $%
\mathbf{\circ }$ & $\mathbf{\circ }$ & $\mathbf{\circ }$ & $\mathbf{\circ }$
& $\mathbf{\circ }$ & $\left\{ 
\begin{array}{c}
\mathbf{\times } \\ 
\circ \\ 
\circ%
\end{array}%
\right. $ & $\left. 
\begin{array}{c}
\mathbf{\circ } \\ 
\times \\ 
\circ%
\end{array}%
\right. $ & $\left. 
\begin{array}{c}
\mathbf{\circ } \\ 
\circ \\ 
\times%
\end{array}%
\right. $ \\ \hline
\multicolumn{1}{|l|}{M5} & $\times $ & $\times $ & $\mathbf{\circ }$ & $%
\mathbf{\circ }$ & $\mathbf{\circ }$ & $\mathbf{\circ }$ & $\times $ & $%
\times $ & $\left\{ 
\begin{array}{c}
\mathbf{\circ } \\ 
\times \\ 
\times%
\end{array}%
\right. $ & $\left. 
\begin{array}{c}
\mathbf{\times } \\ 
\circ \\ 
\times%
\end{array}%
\right. $ & $\left. 
\begin{array}{c}
\mathbf{\times } \\ 
\times \\ 
\circ%
\end{array}%
\right. $ \\ \hline
\end{tabular}%
$%
\end{tabular}%
\end{equation}%
In the case where the 1-forms $\Delta _{i}\neq 0$ and $\lambda _{\alpha
}\neq 0$ and the other vanishing, the attractor eqs reduce to%
\begin{equation}
\begin{tabular}{lll}
$\left( Y^{a}\mathcal{H}_{ab}^{i}Y^{b}-\tilde{Y}^{a}\mathcal{H}_{ab}^{i}%
\tilde{Y}^{b}\right) $ & $=0$ & , \\ 
$\sum\limits_{i,j,r}\left( Z_{r}^{i}\tau _{ij}^{0}Z_{r}^{j}\right) $ & $=0$
& ,%
\end{tabular}%
\end{equation}%
lead to $Y^{b}=\pm \tilde{Y}^{a}$ and $Z_{el,r}=\pm Z_{mag,r}$.

\subsection{Geometries with AdS$_{2}$}

In this subsection, we study explicitly the attractor mechanism for three
examples of near horizon geometries with non compact AdS$_{2}$ factors;
these are: (a) \emph{AdS}$_{2}\times $\emph{S}$^{3}\times $\emph{T}$^{3}$,
(b) \emph{AdS}$_{2}\times $\emph{S}$^{2}\times $\emph{T}$^{4}$ and (c) \emph{%
AdS}$_{2}\times $\emph{S}$^{4}\times $\emph{T}$^{2}$

\subsubsection{\emph{AdS}$_{2}\times $\emph{S}$^{3}\times $\emph{T}$^{3}$}

On this geometry, the non vanishing field strength charges are as follows%
\begin{eqnarray}
&&%
\begin{tabular}{l|l|l|}
\cline{2-3}
& p-branes & $\left( 4-p\right) $- branes \\ \hline
\multicolumn{1}{|l|}{$p=0$} & $\mathcal{F}_{2}^{ai}=Q^{ai}$ $\alpha _{_{AdS_{%
{\small 2}}}}$ & $\mathcal{\tilde{F}}_{6|ai}=P_{ai}$ $\beta _{S^{3}}\wedge
\beta _{T^{3}}$ \\ \hline
\multicolumn{1}{|l|}{$p=1$} & $\mathcal{F}_{3}^{a}=\sum\limits_{r=1}^{2}%
\text{ }p^{a}\beta _{S^{3}}$ & $\mathcal{\tilde{F}}_{5|a}=\sum%
\limits_{r}q_{a}\alpha _{_{AdS_{{\small 2}}}}{\small \wedge }$ $\beta
_{T^{3}}$ \\ \hline
\multicolumn{1}{|l|}{$p=2$} & \multicolumn{2}{|l|}{$\ \ \ \ \mathcal{F}%
_{4}=\sum e_{I}$ $\varepsilon ^{IJK}\alpha _{_{AdS_{{\small 2}}}}{\small %
\wedge }\beta _{S_{J}^{1}}\wedge \beta _{S_{K}^{1}}+\sum g^{I}\beta
_{S^{3}}\wedge \beta _{S_{I}^{1}}$} \\ \hline
\end{tabular}
\\
&&  \notag
\end{eqnarray}%
where $I,J,K=1,2,3$ are associated with the three 1-cycle generators of the
3-torus. From these field strengths, we learn that this geometry supports:%
\newline
(\textbf{i}) $3\times 2$ electrically charge black holes with charges $%
Q^{ai} $ and their magnetic duals with magnetic charges $P_{ai}$,\newline
(\textbf{ii}) three magnetically charged strings with charge $p^{a}$, and
three electrically charged 3-branes with charge $q_{a}$,\newline
(\textbf{iii}) three dyonic membranes with charges $\left(
g^{I},e_{I}\right) .$\newline
Following the same approach we have been using, the effective potential $%
\mathcal{V}_{eff}$ of these black brane configurations reads as follows,%
\begin{equation}
\begin{tabular}{ll}
$\mathcal{V}_{eff}=$ & $\frac{1}{2}\sum \left( X^{ai}\delta _{ab}\delta
_{ij}X^{bj}+\tilde{X}_{ai}\delta ^{ab}\delta ^{ij}\tilde{X}_{bj}\right) $ \\ 
& $+\frac{1}{2}\sum \left( Y^{a}\delta _{ab}Y^{b}+\tilde{Y}_{a}\delta ^{ab}%
\tilde{Y}_{b}\right) $ \\ 
& $+\frac{1}{2}\sum Z_{I}^{i}\delta _{ij}Z_{I}^{j}$%
\end{tabular}%
\end{equation}%
where summation over the various indices is understood. The extremization of
this effective potential leads to%
\begin{equation}
\delta \mathcal{V}_{eff}=\sum\limits_{A}\Delta _{A}\Upsilon
^{A}+\sum\limits_{\alpha }\lambda _{\alpha }\digamma ^{\alpha }=0,
\end{equation}%
with 
\begin{equation}
\begin{tabular}{ll}
$\Upsilon ^{A}=$ & $\left( Y^{a}\mathcal{T}_{ab}^{A}Y^{b}-\tilde{Y}^{a}%
\mathcal{T}_{ab}^{A}\tilde{Y}^{b}\right) +\sum\limits_{i=1}^{2}\left( X^{ai}%
\mathcal{T}_{ab}^{A}X^{bi}-\tilde{X}^{ai}\mathcal{T}_{ab}^{A}\tilde{X}%
^{bi}\right) $ \\ 
$\digamma ^{\alpha }=$ & $\left[ \sum\limits_{a=1}^{3}\left( X^{ai}\tau
_{ij}^{\alpha }X^{aj}-\tilde{X}^{ai}\tau _{ij}^{\alpha }\tilde{X}%
^{bj}\right) +\sum\limits_{I}Z_{I}^{i}\tau _{ij}^{\alpha }Z_{I}^{i}\right] $%
\end{tabular}%
\end{equation}%
The solutions of these attractor eqs $\Upsilon ^{A}=0,$\ $\digamma ^{\alpha
}=0$ may be realized in various ways; one of them is given by the following:%
\begin{equation}
\begin{tabular}{lllll}
$Y^{a}=\pm \tilde{Y}^{a}$ & $,$ & $X^{ai}=\pm \tilde{X}^{ai}$ & , & $%
Z_{I}^{i}=0$%
\end{tabular}%
\end{equation}%
The solutions with plus signs describe intersecting attractor involving 
\emph{three} strings, \emph{three} 3-branes, \emph{six }black holes\emph{\ }%
and \emph{six }4- branes; but no membrane; those with minus signs are
associated with the corresponding anti-branes.

\subsubsection{\emph{AdS}$_{2}\times $\emph{S}$^{2}\times $\emph{T}$^{4}$}

On this geometry, the general form of the field strengths reads as follows,%
\begin{eqnarray}
&&%
\begin{tabular}{|l|l|}
\cline{1-2}
p-branes & $\left( 4-p\right) $- branes \\ \hline
$\mathcal{F}_{2}^{ai}=P^{ia}\beta _{S^{2}}$ & $\mathcal{\tilde{F}}%
_{6|ai}=Q^{ia}\alpha _{Ads_{2}}\wedge \beta _{T^{4}}$ \\ \hline
$\mathcal{F}_{3}^{a}=\sum p^{ak}\left( \beta _{S^{2}}\wedge \beta
_{S_{k}^{1}}\right) $ & $\mathcal{\tilde{F}}_{5|a}=\sum q_{ai}\varepsilon
^{ijkl}\left( \alpha _{Ads_{2}}\wedge \beta _{S_{j}^{1}}\wedge \beta
_{S_{k}^{1}}\wedge \beta _{S_{l}^{1}}\right) $ \\ \hline
\multicolumn{2}{|l|}{$\ \ \ \ \mathcal{F}_{4}=\sum \mathbf{e}_{\left[ kl%
\right] }\varepsilon ^{klrs}\left( \alpha _{Ads_{2}}\wedge \beta
_{S_{r}^{1}\times S_{s}^{1}}\right) +g^{\left[ rs\right] }\left( \beta
_{S^{2}}\wedge \beta _{S_{r}^{1}\times S_{s}^{1}}\right) $} \\ \hline
\end{tabular}
\\
&&  \notag
\end{eqnarray}
The total effective potential reads, in terms of the dressed central charges
of the black holes/4-branes $\left( X^{ia},\tilde{X}_{ia}\right) $, the 
\emph{four} triplets of black strings/3-branes $\left( Y_{k}^{a},\tilde{Y}%
_{a}^{k}\right) _{1\leq k\leq 4}$ and the \emph{six-uplet} dressed electric
charges $Z_{\left[ kl\right] }^{elc}=Z_{\left[ kl\right] }^{1}$ and magnetic 
$Z_{\left[ kl\right] }^{mag}=Z_{\left[ kl\right] }^{2}$ of the dyonic
membranes, as follows: 
\begin{equation}
\begin{tabular}{ll}
$\mathcal{V}_{eff}=$ & $\frac{1}{2}\sum\limits_{a,b=1}^{3}\sum%
\limits_{i,j=1}^{2}\left( X_{I}^{ai}\delta _{ab}\delta _{ij}X_{I}^{bj}+%
\tilde{X}_{ai}^{I}\delta ^{ab}\delta ^{ij}\tilde{X}_{bj}^{I}\right) $ \\ 
& $+\frac{1}{2}\sum\limits_{k=1}^{4}\sum\limits_{a,b=1}^{3}\left(
Y_{k}^{a}\delta _{ab}Y_{k}^{b}+\tilde{Y}^{aj}\delta _{ab}\tilde{Y}%
^{bj}\right) $ \\ 
& $\frac{1}{2}\sum\limits_{k,l=1}^{4}\sum\limits_{i,j=1}^{2}\left( Z_{\left[
kl\right] }^{i}\delta _{ij}Z_{\left[ kl\right] }^{j}\right) $%
\end{tabular}%
\end{equation}%
Its gives the attractor eqs%
\begin{equation}
\begin{tabular}{lll}
$Tr\left( X\mathcal{T}^{A}X\right) -Tr\left( \tilde{X}\mathcal{T}^{A}\tilde{X%
}\right) $ & $=0$ & , \\ 
$\sum\limits_{k=1}^{4}\left[ Tr\left( Y_{k}\mathcal{T}^{A}Y_{k}\right)
-Tr\left( \tilde{Y}_{k}\mathcal{T}^{A}\tilde{Y}_{k}\right) \right] $ & $=0$
& , \\ 
$\sum\limits_{k,l=1}^{4}Tr\left( Z_{\left[ kl\right] }\tau ^{\alpha }Z_{%
\left[ kl\right] }\right) $ & $=0$ & .%
\end{tabular}%
\end{equation}%
The two first relations are solved as usual; i.e $X^{ai}=\pm \tilde{X}^{ai}$%
, $Y_{k}^{a}=\pm \tilde{Y}_{k}^{a}$ while the third has various solution
based on choices that lead to $Tr\left( \tau ^{\alpha }\right) $. These
solutions corresponds to diverse configurations involving intersecting of
black hole, black 4-brane, black string, black 3-brane and black 2-brane.

\subsubsection{\emph{AdS}$_{2}\times $\emph{S}$^{4}\times $\emph{T}$^{2}$}

Using the various n-cycles of \emph{AdS}$_{2}\times $\emph{S}$^{4}\times $%
\emph{T}$^{2}$ and the corresponding n-forms that could live on, the general
expressions of the field strengths on this geometry reads as follows, 
\begin{equation}
\begin{tabular}{|l|l|}
\hline
p-branes & $\left( 4-p\right) $- branes \\ \hline
$\mathcal{F}_{2}^{ai}=Q^{ia}\alpha _{AdS_{2}}$ & $\mathcal{\tilde{F}}%
_{6|ai}=P_{ia}$ $\beta _{\emph{S}^{4}}\wedge \beta _{\emph{T}^{2}}$ \\ \hline
$\mathcal{F}_{3}^{a}=\sum\limits_{k=1}^{2}q_{k}^{a}\left( \alpha
_{AdS_{2}}\wedge \alpha _{S_{k}^{1}}\right) $ & $\mathcal{\tilde{F}}%
_{5|a}=\sum\limits_{k=1}^{2}p_{a}^{k}\left( \beta _{\emph{S}^{4}}\wedge
\alpha _{S_{k}^{1}}\right) $ \\ \hline
\multicolumn{2}{|l|}{$\ \ \ \ \ \ \ \ \mathcal{F}_{4}=\mathbf{e}\left(
\alpha _{AdS_{2}}\wedge \alpha _{T^{2}}\right) +g\beta _{S^{4}}$} \\ \hline
\end{tabular}%
\end{equation}%
where now the strings are charged electrically and the 3-branes
magnetically. The total effective potential $\mathcal{V}_{eff}$ associated
with this system is given as usual by the sum of the contribution of each
extremal black-brane. The attractor equations following from the
extremization of $\mathcal{V}_{eff}$ are then given by:%
\begin{equation}
\begin{tabular}{lll}
$Tr\tilde{X}\mathcal{T}^{A}\tilde{X}-Tr\left( X\mathcal{T}^{A}X\right)
+\sum\limits_{k=1}^{2}\left[ Tr\left( \tilde{Y}_{k}\mathcal{T}^{A}\tilde{Y}%
_{k}\right) -Tr\left( Y_{k}\mathcal{T}^{A}Y_{k}\right) \right] $ & $=0$ & ,
\\ 
$Tr\left( Z\tau ^{\alpha }Z\right) +Tr\left( X\mathcal{T}^{A}X\right)
-Tr\left( \tilde{X}\tau ^{\alpha }\tilde{X}\right) $ & $=0$ & ,%
\end{tabular}%
\end{equation}%
whose solutions are given by $\tilde{Y}_{k}^{a}=\pm Y_{k}^{a}$, $X^{ia}=\pm 
\tilde{X}^{ia},$ $Z^{i}=0$.

\section{Conclusion}

Motivated by the new results obtained in \cite{8A}, we have focused in this
paper on \emph{8D} maximal supergravity embedded in \emph{11D} M-theory on $%
T^{3}$; and studied the attractor mechanism of black p-branes and their
intersections. In particular, we have considered different configurations of
black brane systems and derived various classes of solutions of their
attractor eqs depending on the values of the dressed charges. \newline
To do so, we first studied the general structure of \emph{8D} non chiral
maximal supersymmetric algebra with p-branes as well its link with M- theory
compactified on $T^{3}$. Then we have developed an unconstrained formalism
to approach the geometry of the moduli space $\left[ SL\left( 3,R\right)
\times SL\left( 2,R\right) \right] /\left[ SO\left( 3\right) \times SO\left(
2\right) \right] $ and the symmetries of effective potential $\mathcal{V}%
_{eff}$ of the black p-branes of the \emph{8D} maximal supergravity. In this
way the scalar moduli of the supergravity are captured by two matrices L$%
_{ab}$ and K$_{ij}$ valued in the $SL\left( 3,R\right) \times SL\left(
2,R\right) $ Lie group manifold; the extra degrees of freedom are suppressed
by requiring gauge invariance under the $SO\left( 3\right) \times SO\left(
2\right) $ isometry. The attractor eqs of the black object of 8D
supergravity have the remarkable factorization, 
\begin{equation}
\begin{tabular}{llll}
$\sum\limits_{A=1}^{8}\Delta _{A}\Upsilon ^{A}=0$ & , & $\sum\limits_{\alpha
=1}^{3}\lambda _{\alpha }\digamma ^{\alpha }=0$ & ,%
\end{tabular}
\label{aaa}
\end{equation}%
with the two terms respectively associated with $\left[ SL\left( 3,R\right)
/SO\left( 3\right) \right] $ and $\left[ SL\left( 2,R\right) /SO\left(
2\right) \right] $; in agreement with the factorized structure of the moduli
space. Using the identities $\Upsilon ^{A}=\Upsilon ^{\left( ab\right) }%
\mathcal{T}_{ab}^{A}$, $\Delta _{A}=\Delta _{\left( ab\right) }\mathcal{T}%
_{A}^{ab}$ and quite similar relations for the $SL\left( 2,R\right) $
factor, these attractor equation may be also put in the equivalent forms $%
\Delta _{ab}\Upsilon ^{ab}=0$, $\lambda _{ij}\digamma ^{ij}=0$. In (\ref{aaa}%
), $\Delta _{A}$, $\lambda _{\alpha }$ (or equivalently $\Delta _{ab},$ $%
\lambda _{ij}$) are 1-forms given by eqs(\ref{fo}) and $\Upsilon ^{A},$ $%
\digamma ^{\alpha }$ (or equivalently $\Upsilon ^{ab}$, $\digamma ^{ij}$)
have the typical expressions, 
\begin{equation}
\begin{tabular}{llll}
$\Upsilon ^{A}=\sum\limits_{I}Tr\left( Y_{I}\mathcal{T}^{A}Y_{I}-\tilde{Y}%
_{I}\mathcal{T}^{A}\tilde{Y}_{I}\right) $ & , & $\digamma ^{\alpha
}=\sum\limits_{r}Tr\left( Z_{r}\mathcal{\tau }^{\alpha }Z_{r}\right) $ & .%
\end{tabular}
\label{bbb}
\end{equation}%
where $Y_{I},$ $\tilde{Y}_{I},$ $Z_{r}$ stand for dressed charges and $%
\mathcal{T}^{A},$ $\mathcal{\tau }^{\alpha }$ for the generators of $%
SL\left( 3,R\right) \otimes SL\left( 2,R\right) $. Similar expression can be
written down for $\Upsilon ^{ab}$ and $\digamma ^{ij}$ leading to $\Upsilon
^{ab}=Y_{I}^{a}Y_{I}^{b}-\tilde{Y}_{I}^{a}\tilde{Y}_{I}^{b}$ and $\digamma
^{ij}=Z_{r}^{i}Z_{r}^{j}$. One of the outcome of these expression is that $%
\Upsilon ^{ab}=\Upsilon ^{ba}$, $\digamma ^{ij}=\digamma ^{ji}$; that is
symmetric tensors \ showing that there is no contribution to the attractor
eqs coming from those generators with $\mathcal{T}_{ab}^{A}=-\mathcal{T}%
_{ba}^{A}$ and $\mathcal{\tau }_{ij}^{\alpha }=-\mathcal{\tau }_{ji}^{\alpha
}$. This property reflects just the decoupling of the contribution
associated with the $SO\left( 3\right) $ factor inside $SL\left( 3,R\right) $%
. A similar conclusion is also valid for the SL$\left( 2,R\right) $
component and the $SO\left( 2\right) $ subroup. \newline
We end this discussion by noting that the solutions worked out in this study
are real solutions since the moduli space $\left[ SL\left( 3,R\right) \times
SL\left( 2,R\right) \right] /\left[ SO\left( 3\right) \times SO\left(
2\right) \right] $ is a real manifold. Complex solutions, such as eq(\ref{12}%
) with $\nu \notin $ $\left[ 0,1\right] $, can be found if instead of (\ref%
{ms}), we use the complex manifold $\left[ SU\left( 1,2\right) /SU\left(
2\right) \right] \times \left[ SU\left( 1,1\right) /U\left( 1\right) \right] 
$.

\begin{acknowledgement}
This research work is supported by the program Protars III D12/25.
\end{acknowledgement}

\appendix

\section{Dirac matrices in \emph{8D}}

Here we give some useful properties on the algebra of $\Gamma $- matrices in
8D space time dimensions. In a quite similar way as in 4D, there are eight $%
\Gamma $- matrices in 8D namely $\Gamma ^{\mu }$ with $\mu =0,...,7$; they
generate the Clifford algebra \textrm{\cite{22A},} 
\begin{equation}
\Gamma ^{\mu }\Gamma ^{\nu }+\Gamma ^{\nu }\Gamma ^{\mu }=2\eta ^{\mu \nu }
\end{equation}%
with $\eta ^{\mu \nu }=diag\left( -,+...+\right) $ standing for the metric
of $R^{1,7}$ with rotation symmetry SO$\left( 1,7\right) $. From these
relations we learn amongst others that $\left( \Gamma ^{0}\right)
^{2}=-I_{id}$ and $\left( \Gamma ^{i}\right) ^{2}=+I_{id}$ for $i=1,...,7$.
The simplest realization of the $\Gamma ^{\mu }$'s is given by $n\times n$
matrices with $n=2^{4}$ and so act on \emph{16} components objects: Dirac
spinors. With these matrices, one can build others carrying several space
time vectors by taking the completely antisymmetric products as follows:%
\begin{equation}
\begin{tabular}{ll}
$\Gamma ^{\mu _{1}...\mu _{p}}=\frac{1}{p!}\left( \Gamma ^{\mu
_{1}}...\Gamma ^{\mu _{p}}\pm \text{{\small permutations}}\right) $ & ,%
\end{tabular}%
\end{equation}%
where $\left( +\right) $ and $\left( -\right) $ stand respectively for even
and odd permutations. For the leading $p=2$ case, we have just the
commutators of $\Gamma $-matrices which give a realization of the $SO\left(
1,7\right) $ rotation generators $M^{\mu \nu }$ in the \emph{16} dimensional
spinor representation,%
\begin{equation}
M^{\mu \nu }=\frac{i}{2}\Gamma ^{\mu \nu }.
\end{equation}%
Moreover and like in 4D, we distinguish three kinds of spinors in \emph{8D};
the first one is the $SO\left( 1,7\right) $ Dirac spinor $\Psi ^{%
{\scriptsize Dirac}}$ having \emph{16} complex components. But this spinor
is reducible into $8+8$ components describing each a complex Weyl spinor
defining the two chiralities of $\Psi ^{{\scriptsize Dirac}}$ namely the
left $\Psi _{L}\equiv \left( \psi \right) _{a}$ and the right $\Psi
_{R}\equiv \left( \chi _{\dot{a}}\right) $ related to the Dirac $\Psi $ as
follows, 
\begin{equation}
\begin{tabular}{llll}
$\Psi _{L}=\frac{1}{2}\left( 1-\Gamma _{8}\right) \Psi $ & , & $\Gamma
_{8}\Psi _{L}=-\Psi _{L}$ & , \\ 
$\Psi _{R}=\frac{1}{2}\left( 1+\Gamma _{8}\right) \Psi $ & , & $\Gamma
_{8}\Psi _{R}=+\Psi _{R}$ & .%
\end{tabular}%
\end{equation}%
In these relations, the matrix $\Gamma _{8}$ is the chirality operator given
by 
\begin{equation}
\Gamma _{8}=e^{-i\frac{3\pi }{2}}\Gamma ^{0}...\Gamma ^{7},
\end{equation}%
satisfying, amongst others, the following properties%
\begin{equation}
\begin{tabular}{llll}
$\left( \Gamma _{8}\right) ^{2}=I,$ & $\left\{ \Gamma _{8},\Gamma ^{\mu
}\right\} =0,$ & $\left[ \Gamma _{8},\Gamma ^{\mu \nu }\right] =0$ & .%
\end{tabular}%
\end{equation}%
The third kind of spinors in \emph{8D} is of Majorana type; that is a Dirac
spinor constrained by the typical reality condition 
\begin{equation}
\begin{tabular}{ll}
$\Psi ^{\ast }=B\Psi $ & ,%
\end{tabular}%
\end{equation}%
where $B$ is $16\times 16$ matrix. This condition which also reads as $\Psi
=B^{\ast }B\Psi $ should be as well consistent with Lorentz transformation $%
\delta \Psi =i\omega _{{\scriptsize \mu \nu }}M^{{\scriptsize \mu \nu }}\Psi 
$. The solution of the constraint relations leads to%
\begin{equation}
\begin{tabular}{llll}
$B^{\ast }B=I$ & , & $M^{\ast \mu \nu }=-BM^{\mu \nu }B^{-1}$ & ,%
\end{tabular}%
\end{equation}%
where $M^{\ast \mu \nu }$ is the Lorentz matrix generating rotation $\Psi
^{\ast }$; that is $\delta \Psi ^{\ast }=-i\omega _{{\scriptsize \mu \nu }%
}\left( M^{\ast {\scriptsize \mu \nu }}\right) \Psi ^{\ast }$.

\end{document}